\documentclass[preprint,12pt]{elsarticle}

\usepackage{graphicx}
\usepackage{hyperref}
\usepackage[table]{xcolor}
\usepackage{floatrow}
\usepackage{multirow}
\usepackage{tabularx}
\usepackage{fontawesome}
\usepackage{amssymb}
\usepackage{lineno}
\usepackage{soul}
\usepackage{subcaption}
\usepackage{hyperref}
\hypersetup{colorlinks=true,urlcolor=blue,citecolor=blue, linkcolor=blue}

\usepackage{natbib}
\setcitestyle{authoryear,open={(},close={)}}

\usepackage{longtable}
\usepackage{array}
\usepackage{tabularx}
\usepackage{graphicx}
\usepackage{adjustbox}

\journal{Journal Name}

\begin{document}

\begin{frontmatter}


\title{Sponsored messaging about climate change on Facebook: Actors, content, frames}



\author[cs]{Iain Weaver}
\author[cs]{Ned Westwood}
\author[pol]{Travis Coan}
\author[geog]{Saffron O'Neill}
\author[cs,turing]{Hywel T.P. Williams\corref{cor1}}
\cortext[cor1]{Corresponding author.}
\address[cs]{Computer Science, University of Exeter, Exeter, EX4 4SB}
\address[pol]{Political Science, University of Exeter, Exeter, EX4 4SB}
\address[geog]{Geography, University of Exeter, Exeter, EX4 4SB}
\address[turing]{Alan Turing Institute, 96 Euston Road, London, NW1 2DB}

\clearpage

\begin{abstract}
Online communication about climate change is central to public discourse around this contested issue. Facebook is a dominant social media platform known to be a major source of information and online influence, yet discussion of climate change on the platform has remained largely unstudied due to difficulties in accessing data. This paper utilises Facebook's recently created repository of social/political advertisements to study how climate change is framed as an issue in adverts placed by different actors. Sponsored content is a strategic investment and presumably intended to be persuasive, so patterns of who pays for adverts and how those adverts frame the issue can reveal large-scale trends in public discourse.
Here we show that most money spent on (English-language) climate-related messaging is targeted at users in the United States, Great Britain and Canada. While the number of advert impressions correlates with total spend by an actor, there is a secondary effect of unpaid social sharing which can substantially affect the number of impressions per dollar spent. Most spend in the US is by political actors, while environmental non-governmental organisations dominate spend in Great Britain; both types of actor are important in Canada, where commercial actors also spend heavily. Content analysis shows that climate change solutions such as tree-planting are well represented in GB, while climate change impacts such as forest fires, flooding and other extreme weather events are strongly represented in the US and Canada. Different actor types frame the issue of climate change in different ways; political actors position the issue as party political and a point of difference between candidates, whereas environmental NGOs frame climate change as the focus of collective action and social mobilisation. Overall, our study provides a first empirical exploration of climate-related advertising on Facebook. It shows the diversity of actors seeking to use Facebook as a platform for their campaigns and how they utilise different topic frames to persuade users to take action.
\end{abstract}

\begin{keyword}
climate change \sep advertising \sep social media \sep campaign


\end{keyword}

\end{frontmatter}

\newpage 
\section{Introduction}
\label{introduction}

Climate change remains a contentious political issue and many groups invest heavily in climate-related messaging. Politicians and political parties embed climate policy in their campaigns. Commercial enterprises incorporate climate and environmental themes into their product advertising. Advocacy groups spread information, endorse or oppose climate policies, and mobilize supporters for action. Increasingly, these groups turn to social media to disseminate their messages, as popular platforms such as Facebook or Twitter offer a relatively inexpensive and effective means to connect with large numbers of potential supporters. These platforms not only offer a framework for user-led dissemination of free content, but are fundamentally commercial enterprises that generate income from targeted advertising and promotion of paid content. The political implications of social media advertising have come under increasing scrutiny by both journalist and lawmakers, following suggestions that online campaigns were influential in (e.g.) the US presidential election and UK `Brexit' referendum in 2016. This increased interest has led to calls for transparency and some social media platforms have begun to open their advertising data for investigation. This study utilises advertising data from the Facebook platform made available by the Facebook Ad Library\footnote{https://www.facebook.com/ads/library/ Date of last access: 19th July 2022}.

Here we study social and political advertising on Facebook related to the contentious issue of climate change. A growing academic literature examines discussion of climate change on social media (e.g. \citep{schafer2012onlinecomms,sharman2014scepticalblogosphere,williams2015echochambers,pearce_social_2019,hautea2021tiktokactivism, vu2021framingNGOfacebook, jang&hart2015twitterframing,treen2022reddit}), but coverage has not been balanced across the many different platforms where such discussions take place. In particular, most studies have focused on Twitter \citep{pearce_social_2019}, most likely due to the relative ease of accessing data from that platform compared to others. Here we present an initial empirical study of climate-related advertising on Facebook, addressing two outstanding research gaps. Firstly, little attention has been devoted to study of climate communication on Facebook, despite it being the most popular social media platform globally \citep{statista2020socialmedia}. Secondly, few (if any) studies focus on climate change advertising on social media. Appropriate to this context, a major aim of the study is to answer basic empirical questions around the phenomenology of climate-related advertising on Facebook: \textit{Who} is advertising and how much do they spend? \textit{What} do adverts contain in terms of content and messaging? \textit{When} does climate advertising appear? Drawing upon the answers to these questions, we also offer a more theoretical contribution around the framing of climate change as an issue in paid Facebook content and what this might mean for public opinion. 

The next section covers the Research Context, expanding on the academic literature relevant to this study and supporting our research questions. This is followed by a brief summary of Methods (further details are given in Supporting Information). Next we present some Exploratory Data Analysis to describe basic properties of the dataset used, followed by more detailed Results relating to our research questions. Finally, the Discussion explores the main findings and implications of the work.

\section{Research Context}

Climate change remains a divisive issue in public discourse, which is increasingly enacted in online social media. The rise of digital media has changed the ways in which social movements emerge and how interest groups, political campaigns, and corporations interact with supporters and other stakeholders \citep{karpf2012moveon,obar2012advocacy2.0,chalmers2016changingface}. Here we review some relevant background material and develop our research questions around study of climate-related advertising on the Facebook platform.

\subsection{Digital advocacy and climate change}

While decades of research exist on environmental advocacy, only recently have scholars turned to systematically examining the importance of digital technologies in this area. Research suggests that online tools in general and social media platforms in particular have significantly reduced the costs of political campaigning \citep{earl2011digital}, increased opportunities for both ``legacy'' advocacy organizations and new ``digitally native'' groups to reach potential supporters \citep{karpf2012moveon}, and thereby promoted new forms of activism \citep{freelon2020science} and civic engagement \citep{obar2012advocacy2.0}. A recent review by \cite{johansson2019digital} categorises digital advocacy studies by their focus on ``digital access politics'' (where the goal is exchange policy-relevant information with government officials), ``digital information politics'' (in which the objective is disseminate information to the general public), or ``digital protest politics'' (where the goal is to promote specific forms of online or offline political activism). Although a number of studies point to the the challenges for effective digital advocacy and meaningful online civic engagement (see \citealt{schmitz2020advocacy}), there is general agreement that organizations must adapt to the information environment if they want to effectively promote their messages \citep{freelon2020science}.

Despite the considerable progress made in understanding online advocacy campaigns across a wide-range of political issues, less attention is devoted specifically to climate change. Several recent studies also examine the activities of advocacy groups on prominent social media platforms around the issue of energy policy. \cite{hodges2016keystone} examine the use of Twitter by environmental non-governmental organizations (ENGOs) in the US to mobilize online and offline action against the Keystone XL pipeline. \cite{kimchi2015greenpeace} analyze Greenpeace's ``Unfriend Coal" campaign on Facebook by combining interviews of Greenpeace staff and a content analysis of posting activity. \cite{hendriks2016dramaturgy} also focus on Facebook, offering an analysis of the competing narratives across supporter and opposition groups around the controversial Narrabri Gas Project in Australia. Meanwhile, \cite{vu2021framingNGOfacebook} studied the framing of climate change by environmental NGOs using Facebook posts on public pages.

\subsection{Climate change advertising}

The use of advertisements to facilitate strategic messaging on environmental issues is not unique to the information age. In 1966, the Sierra Club ran full page advertisements in the \textit{New York Times} and \textit{Washington Post} to raise awareness on the United States government's plans to build dams in National Forests in the American west, warning readers that ``[N]ow Only You Can Save Grand Canyon From Being Flooded'' \citep{wyss2016sierra}. Energy companies, moreover, have a long history of implementing advertising campaigns on environmental issues, as demonstrated recently by the `Clean Coal' campaign in the United States, the United Kingdom, and Canada \citep{fitzgerald2012cleancoal}. There now exists a well-established literature on `green advertising' \citep{banerjee1995greenads, hartmann2009greenads} and an active debate on the influence of environmental messages on consumer behavior (see \citealt{xue2015greenads}), as well as the role of advertising in addressing the `green gap' between what individuals feel they should to to protect the environment and what they are actually doing. 

While the literature on green advertising is extensive, less work focuses specifically on climate change advertising. Much scholarship views climate advertising through the lens of social marketing, where the objective of advertisers is to promote behavior change that supports the social good \citep{peattie2009climateads, corner2011climateads}. Within the context of social marketing, scholars tend to focus on energy consumption, the transition to renewable forms of energy, and promoting the use of public transportation \citep{takashi2009socialmarketing}. Some researchers remain doubtful of the ability of social marketing campaigns alone to promote the significant changes to behavior necessary to avoid the worst effects of climate change \citep{cornerrandall2015socialmarketing}. Environment-related advertising can also be targeted against pro-environment actions or beliefs. \cite{supran2017assessing} document ExxonMobil's use of `advertorials' (i.e., editorial-style advertisements) in casting doubt on the issue of climate change. The ability of these `counter-frames' to confuse debate and obscure accurate climate information is well-documented in psychological research on misinformation \citep{cooketal2017misinformationinoculation, lewandowskyetal2012misinformation} and demonstrates further the challenges of ``selling climate change" \citep{corner2011climateads}. 

Researchers investigating green advertising on social media have more hope for behavioural change when incorporating the effects of group and bandwagon cues into their analytical framework \citep{pittmanreadchen2021greensocialmedia}. The bandwagon heuristic describes the effect of choices made by other users in online communication platforms on the choices of the individual \citep{sundar2008mainmodel}. \cite{pittmanreadchen2021greensocialmedia} find that non-green consumers are motivated towards green purchasing behaviours on social media platforms when presented with low-information/high-fear advertisements, due to increased levels of self-consciousness when compared to non-social website advertisement.

\subsection{Online communication about climate change}

Traditional media have long been studied for their role in shaping public opinion about climate change, but online studies only began to emerge in the last decade as social media grew in usage and importance (e.g. \cite{schafer2012onlinecomms, auer2014microblogs}). Social media differ from traditional media (broadcast, print) in that they are decentralised and participatory, allowing study of social interactions and social network structures alongside the media content itself. 
Early studies related to climate change focused on social media discourse around the IPCC Fifth Assessment Report \citep{pearce2014twitter, oneill2015dominant}, at linkages between climate sceptic bloggers \citep{sharman2014scepticalblogosphere, elgesem2015blogosphere}, at global activity and discussion topics on Twitter \citep{kirilenkoandstepchenkova2014microblogging} and on the social network structures of online discussion \citep{williams2015echochambers}. 

A major focus of climate-related study of social media has been on ideological polarisation and segregation of differing views on the topic \citep{williams2015echochambers, cannweaverwilliams2021biases}. Building on earlier work that studied online polarisation and echo chambers in US politics \citep{adamicandglance2005dividedtheyblog, sunstein2009republic.com2.0, conover2011polarization, conover2012partisan, bakshy2015exposure}, similar methods were applied to show polarisation affecting online climate discourse, with both environmentalist and climate denial perspectives being put forward in largely isolated echo chambers \citep{elgesem2015blogosphere, williams2015echochambers, walkterandbruggemannengesser2018echochambers, cannweaverwilliams2021biases}. Such divisions arise out of a long history of contrarian argumentation by climate `deniers' or `sceptics' \citep{hulme2009controversy}, which has been updated for the digital age with novel aspects such as online bots, artificial amplification and platform manipulation \citep{treen2020misinfo}. The different constituencies in the online debate have also been studied from the perspectives of content (e.g. \cite{boussalisandcoan2016textmining}) and framing (e.g. \cite{jang&hart2015twitterframing,vu2021framingNGOfacebook}). Despite the overwhelming scientific consensus on the human-made impacts of climate change (IPCC), the active and visible denial lobby has continued to create doubt and slow political progress towards meaningful solutions \citep{farrell2018misinformation,treen2020misinfo}. However, there is a growing focus on countering climate misinformation, with various theories and tools being put forward to detect misinformation and lessen its effects (e.g. \cite{treen2020misinfo,cooketal2017misinformationinoculation, vanderlinden2017inoculatingpublic, farrell2018misinformation, coan2021computer}).

There remain many research gaps in our understanding of climate discourse in online media, compounded by the rapidity of change in the underlying technologies and platforms, as well as changing behaviours and usage patterns. It is clear that online media is now central in shaping public opinion and requires interdisciplinary study. One weakness identified in the literature has been an over-dependence on Twitter as a data source (Pearce et al, 2019). While Twitter is an important platform for climate discourse, used widely by a variety of actors on all sides of the debate, its status as the most-studied platform arises partly from the relative ease of accessing data (contrast the restrictive data-sharing situation for other platforms such as Facebook, Instagram and WhatsApp \citep{pearce_social_2019}). There is a clear need for more breadth in our understanding of social media discussion of climate change, notwithstanding recent work that is beginning to broaden the coverage \citep{treen2022reddit}.

\subsection{Framing climate change}

Framing theory is often used as a conceptual lens with which to understand the strategic communication of actors involved in climate discourse. `Proximity framing' represents efforts to lengthen or shorten psychological distances between the individual and climate change. For example, \cite{wiest2015partisanframing} test for the effects of local vs global frames with regards to climate impacts, finding local frames more able to influence opinions towards action on climate change. `Issue framing' represents efforts to drive a policy issue in a certain narrative direction by association with other policy issues and topics. \cite{hartfeldman2014threatefficacy} use issue framing in an analysis of US news coverage, finding that climate impacts and climate action were often discussed in separate broadcasts, with the former being discussed under environmental frames (i.e. ecological changes) and the latter highlighting the political struggles associated with action. Other perspectives include `collective action frames', popularised by \cite{benfordsnow2000framing} to reveal framing activities vital to social movement organisations, `visual frames' \citep{oneill2013imagematters} and `equivalency frames' \citep{druckman2004framingeffects}, among others.

Whilst much attention has been given to news broadcasts and print media, a growing number of studies focus generally on how climate change is framed on social media sites such as Twitter \citep{jang&hart2015twitterframing}. Particular emphasis has been given to the framing of international events and climate reports \citep{oneill2015dominant, newman2017twitter, kimcook2018trumpframing}, the prevalence of themes discussed in particular user communities \citep{pearce2014twitter,williams2015echochambers}, differences between different national contexts \citep{jang&hart2015twitterframing}, and the effects of extreme weather patterns \citep{moernaut2020hotweatherframing}. Whilst a growing literature exists for Twitter, less attention has been given to other social media sites such as Facebook, Instagram, TikTok and YouTube \citep{pearce_social_2019}. \cite{vu2021framingNGOfacebook} conduct an analysis of public Facebook posts by NGOs, utilising \cite{benfordsnow2000framing}'s framing schema and building on work from other scholars that have tested for framing effects of impacts, actions and efficacy messages \citep{hartfeldman2014threatefficacy, smithjoffe2009visualframing, oneillnicholsoncole2009positiveframing}. \cite{hautea2021tiktokactivism} conduct an analysis of TikTok videos under the climate change hashtag. \cite{pearce_social_2019} highlight the scholarly focus on Twitter and suggest future research consider other platforms, as well as broader consideration of non-textual elements such as images, which ``play a pivotal role within the richly multimodal nature of many platform vernaculars''. 

Visual framing has always played an important role in climate communications. In an analysis of US, UK and Australian newspapers, \cite{oneill2013imagematters} identified popular themes in newspaper climate imagery, showing that identifiable people, particularly politicians, are dominant. Similar results are found by \cite{difrancescoyoung2011visualconstruction} in an analysis of Canadian news papers and \cite{wozniak2017visualframingcontests} in an analysis of COP media coverage. The proliferation of digital media reporting and participatory engagement has further highlighted the importance of visual framing for climate communications. A related research gap also exists in understanding how images and text combine to frame an issue. Multimodal text and image analysis has been highlighted as a key area for future research climate communication scholars \citep{feldmanhart2018framingpolarizingcue}.


\subsection{Research questions \& scope}

Digital advocacy is an important aspect of climate debate and online advertising is expected to be a key tool in the activities of many groups seeking to influence public opinion. Advocacy and advertising form part of a wider online discourse around climate change, in which debate is increasingly enacted through social media platforms. Many academic studies have documented different aspects of online climate debate, such as the fragmentation into polarised echo chambers, the prevalence of misinformation and how the climate issue is framed in different media channels. However, studies of online communications have been mostly focused on Twitter due to easy access to data. Facebook is a major platform worldwide, with a large user population and massive revenue generated by selling targeted advertising. With the creation of the Facebook Ad Library, it is now possible to study how different actors use paid content and attempt to influence public opinion via the Facebook platform. 

Here we address two main research questions: (1) Which actors advertise on Facebook with regard to climate change? (2) What imagery and text content is used in climate-related adverts and how does it frame the issue of climate change?

In this study we focus primarily on English-language content from the USA, UK and Canada. This choice is partly driven by the data available from the Facebook Ad Library archive, which has better coverage for the US and UK than for other countries, but also by theoretical interest. In terms of climate policy, the UK and USA have both been influential. The UK showed global leadership with its Climate Change Act 2008, setting out the first legally binding national mitigation target. In the USA, federal climate policy is beset by party-political struggle; the USA has recently withdrawn (under the Trump administration) and rejoined (under the Biden administration) the Paris Agreement. Both the UK and USA have some of the highest rates of internet penetration globally (95\% and 96\% respectively \citep{newman2021digital}), with Facebook one of the most popular social media platforms (65\% of internet users use Facebook in the UK, 58\% in the USA \citep{newman2021digital}). These factors make the US and UK salient cases for examining how advertising shapes the cultural politics of climate change. Canada also has interesting political factors and is found here to be the third-largest market for climate-related advertising on Facebook (see below), so acts as a useful comparator example for the US and UK.

\section{Methods}
\label{method}

To address our research questions, we first collected sponsored posts (adverts) from the Facebook Ad Library. Then we perform exploratory data analysis to understand the prevalence of climate-related adverts by geography and over time, as well as the important relationship between spend and reach. Next we focus on the key actors in terms of spend and reach, focusing on the US, UK and Canada. Then for a sample of the most-seen adverts, we investigate the images and textual content that form the adverts, to identify how the issue of climate change is framed by different actors. 

Throughout the next sections, we give the main  aspects of our methods close to the results that draw upon them. Detailed methods are given in Supporting Information (SI) and referenced appropriately from the main text.

\section{Exploratory Data Analysis}

\subsection{Data collection}

The Facebook Ad Library makes available an archive of all content that owners of public Facebook Pages have paid to promote to users, with the significant caveat that the Ad Library will only return those posts which have been categorised by Facebook as relating to social issues, elections or politics \citep{fbManual}. While the terminology of the Facebook platform refers to `adverts', it is more general to see them as sponsored posts; the posts are visually similar to other posts (e.g. social messages from friends) except for a small disclaimer message (see examples in Figure \ref{fig:top_ad_grid}) and appear alongside other posts in a user's feed. 

Sponsored posts were retrieved from the Facebook Ad Library API using a keyword search on `climate' and selecting all posts created before 1st April 2020. This resulted in 215,520 unique sponsored posts (adverts) with an estimated total spend value of 45.6m USD; the average spend per advert is approximately 212 USD. See Supporting Information for further details of data extraction (Section~S1.1) and how spend value is calculated for each sponsored post (Section~S1.2).

\subsection{Examples of sponsored posts}

\begin{figure}
    \centering
    \includegraphics[height=.9\textheight]{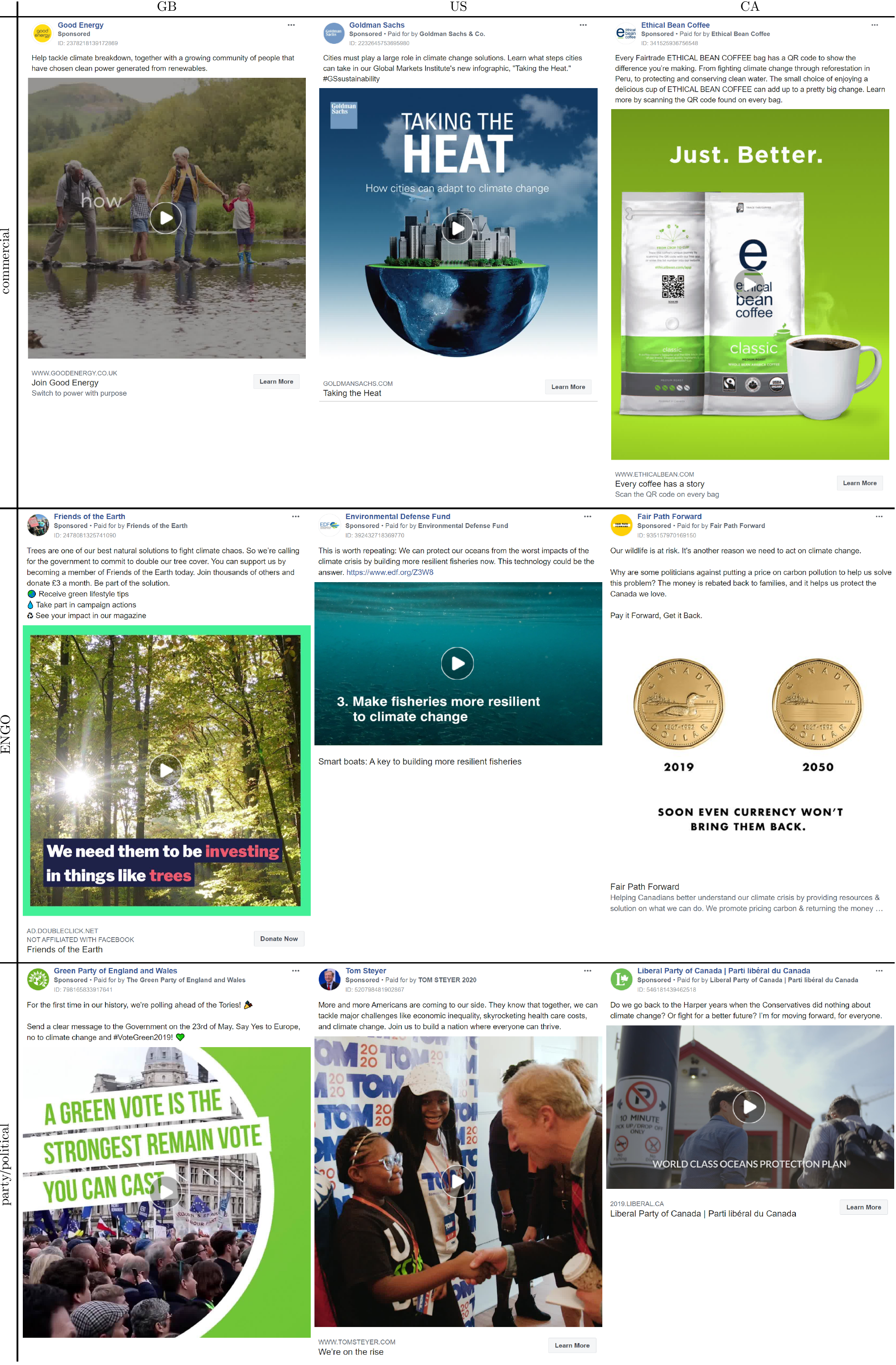}
    \caption{Example sponsored posts relating to climate change. These posts are the most-seen (highest impressions) in each of three countries (US, UK, Canada) originating from three types of actor (environmental non-government organisations (ENGO), commercial organisation, political party or candidate). Posts are shown as they would appear on a user's web browser. The \faPlayCircle~icon indicates that a displayed image is the poster for an embedded video; all the examples contained a video, but many adverts contain only a still image.}
    \label{fig:top_ad_grid}
\end{figure}

Before moving further, it is useful to look at some examples of sponsored posts. Figure \ref{fig:top_ad_grid} shows three sponsored posts from each of three countries (US, UK and Canada). The examples show the most-seen post in each country from each of three types of actor (political, commercial, environmental non-governmental organisation). Looking at the posts, they are all reasonably similar in layout, which is largely imposed by the Facebook platform. At the top of each post, the name of the Facebook Page from which it originates is given, followed by the word ``Sponsored'' to indicate that the post is paid content (and not, for example, a social post by a friend). Then follows the body of the post, almost always containing an image or still frame from a video, with a small amount of text. At the foot of the advert is the name of the sponsor organisation, sometimes with a clickable call to action (e.g. ``learn more'', ``donate now'').

\subsection{Geographic distribution of sponsored posts}

\begin{figure}
\centering
\includegraphics[width=.8\linewidth]{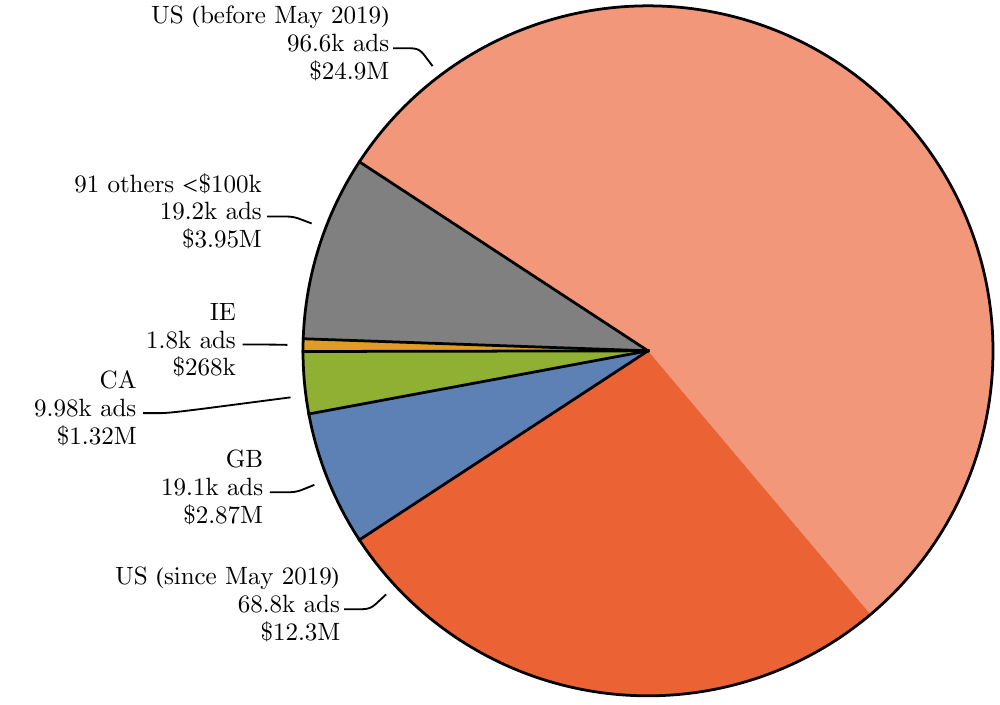}
\caption{Estimated total spend on sponsored Facebook posts using the keyword `climate' by country (in USD). The Facebook Ad Library does not return posts for countries outside the US prior to May 2019; the US total is split into portions before and after this point to enable comparison. Note that the English-language query will have biased the search towards English-speaking countries.}
\label{fig:spend-by-country}
\end{figure}

Figure \ref{fig:spend-by-country} shows the total estimated spend in USD by country. The USA had the largest number of posts when searching by keyword `climate', followed by the UK and Canada. Note that the English-language query is likely to have favoured English-speaking countries, which dominate the dataset. Figure \ref{fig:spend-by-country} partitions the US posts into portions before and after May 2019 to enable fair comparison to other countries; from our collection, it appears that at time of collection the Facebook Ad Library API did not report sponsored posts from non-US regions before this date. 
Even considering only sponsored posts delivered after this point, the United States (US) clearly dominated Facebook advertising spend for the keyword `climate', followed by the United Kingdom (GB) and Canada (CA), which both spent several times more than the next highest spender, Ireland (IE). 
From this point onward, we focus our investigation onto the three countries with highest spend: the United States, Great Britain and Canada. Considering the spend per capita in these three countries, values are broadly similar: approximately 0.04 USD per capita in the USA, 0.04 USD per capita for the UK, 0.03 USD per capita for Canada\footnote{Populations for 2020 taken from the World Bank Data Catalog `World Development Indicators' dataset https://datacatalog.worldbank.org/search/dataset/0037712. Last access: 31st July 2022.}.

\subsection{Common phrases in climate-related posts}

To better understand the phrasing associated with climate-related posts, and to focus the data more precisely on climate change, we counted all bigrams (two-word phrases) containing the word `climate' from the text fields of all the posts. The top-ranked bigrams were inspected for irrelevant usage (e.g. `political climate') and posts from US, GB and CA were filtered to retain only those containing at least one of the 52 top-ranked bigram phrases related to climate change (see Section S1.3 and Table S3). This filter retains 80\% of posts (equivalent to 82\% of total spend) from the original dataset and ensures that retained posts are unambiguously referring to climate change. 

The top-ten relevant bigrams by volume are: `climate change' (175,808 posts), `climate action' (34,525 posts), `climate crisis' (32,892 posts), `climate emergency' (11,648 posts), `climate future' (8,630 posts), `climate reality' (7,231 posts), `climate education' (4,001 posts), `climate denier' (3,632 posts), `global climate' (3,599 posts), `climate chaos' (3,574 posts). Bigram usage varies between geographies (Figure S2); while `climate change' is the most common bigram for US, GB and CA, the second-ranked bigram varies geographically (`climate crisis' in US and CA, `climate emergency' in GB). Bigrams also overlap, with many posts containing multiple relevant bigrams.

\subsection{Temporal distribution of sponsored posts}

\begin{figure}
    \centering
    \includegraphics[width=0.95\textwidth]{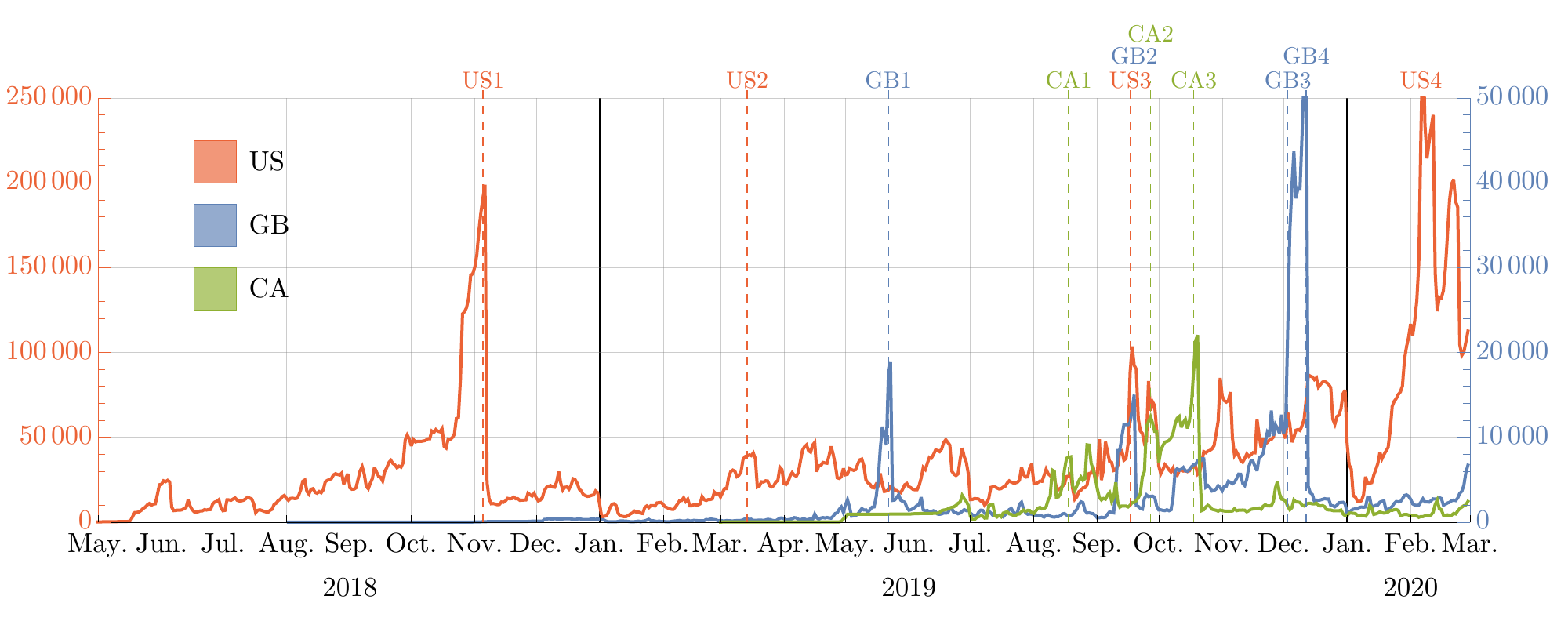}
    \caption{Total daily spend on sponsored posts in USD for United States (US), Great Britain (GB) and Canada (CA). {\em Left axis:} US expenditure. {\em Right axis:} GB and CA expenditure. Significant spikes in activity are mostly correlated with important political events, and occasionally, with extreme weather.}
    \label{fig:concurrent_adverts}
\end{figure}

Looking at the volumes of adverts placed over the time period of this study, it is clear that spending varies substantially over time and appears to be correlated with external events. Figure \ref{fig:concurrent_adverts} shows the total daily expenditure on concurrently running climate-related posts for the US, GB and CA. Daily expenditure was calculated by first calculating the average daily spend for each post (i.e. total post spend divided by duration) and then summing to give the total daily spend across all sponsored posts being delivered on each day in each location. Timeseries for all three countries are characterised by a number of peaks in the total value of posts, typically with a gradual build-up to each peak followed by a sudden drop-off as many posts simultaneously cease delivery. This suggests that sponsored posts are released in the build-up to a particular event.

Manual inspection of post content and reference to external sources (mainly the Wikipedia world news timeline\footnote{Wikipedia Current Events, \url{https://en.wikipedia.org/wiki/Portal:Current\_events} accessed 07-10-2020}) suggests that these peaks are mostly correlated with political events, and occasionally with major weather events, in each country. For example, for the US there are notable peaks on: 5th November 2018 (US1 - the day before the 2018 mid-term elections); 14th March 2019 (US2 - Beto O'Rourke announces his presidential campaign); 17th September 2019 (US3 - Donald Trump impeachment hearings begin, tropical storm Imelda makes landfall in Texas); and 5th February 2020 (US4 - US senate votes on Trump impeachment). For the UK, there are notable peaks for: 23rd May 2019 (GB1 - European Parliament elections); 18th September 2019 (GB2 - worldwide strikes preceding the UN climate summit); and a double peak spanning 3rd December 2019 (GB3 - COP25 climate policy negotiations in Madrid) and 12th December 2019 (GB4 - UK General Election). For Canada, there are peaks around: 18th August 2019 (CA1 - no obvious national interest, but global media interest in Hong Kong protests); 27th September 2019 (CA2 - climate strikes in Montreal); and 21st October 2019 (CA3 - federal election). The shape of the peaks, with a relatively gradual increase to the peak followed by an abrupt drop post-event, is consistent with a pattern of increasing spending in the run-up to a political event and immediate cessation once the event has occurred.

\subsection{Spend and impressions}

Next we study the relationship between the amount spent sponsoring a post and the visibility it gains. The Facebook Ad Library API provides the number of impressions for each post as a measurement of visibility/reach; an impression is simply an instance of the post appearing on a user's feed. It does not imply that the user interacted with the content, only that it was seen (or at least, it was presented in the user's feed while the user was logged in and the feed was displayed on screen). Impressions can be gained in two ways. Firstly, the post can be placed in user feeds algorithmically by the platform, in return for payment by the post sponsor; in this case, the money spent on an advert guarantees a certain number of impressions. Secondly, a sponsored post seen by a user may be voluntarily shared with their Facebook friends, generating more impressions without additional investment by the sponsor. Such social sharing is clearly desirable for sponsors but can be hard to achieve (cf. `viral advertising'). The API does not distinguish between impressions gained in different ways, though the amount of social sharing can be estimated from the volume of impressions per unit spend compared across different posts. 

Here we analyse spend and impression statistics for adverts from the US dataset; similar trends were observed for the CA and GB datasets (data not shown). For this analysis, we aggregate spend and impressions over time for each advert sponsor and count the number of sponsors that fall into different total-spend-vs-total-impressions bands. Analysis of spend over time for each advert shows that most adverts have a total spend in the range 10-1000 USD and a lifetime in the range 1-100 days (Figure S1). Further detail on how spend and impressions are calculated for each advert is given in Section S1.2.

\begin{figure}
    \centering
    \subfloat[][]{\includegraphics[width=.49\textwidth]{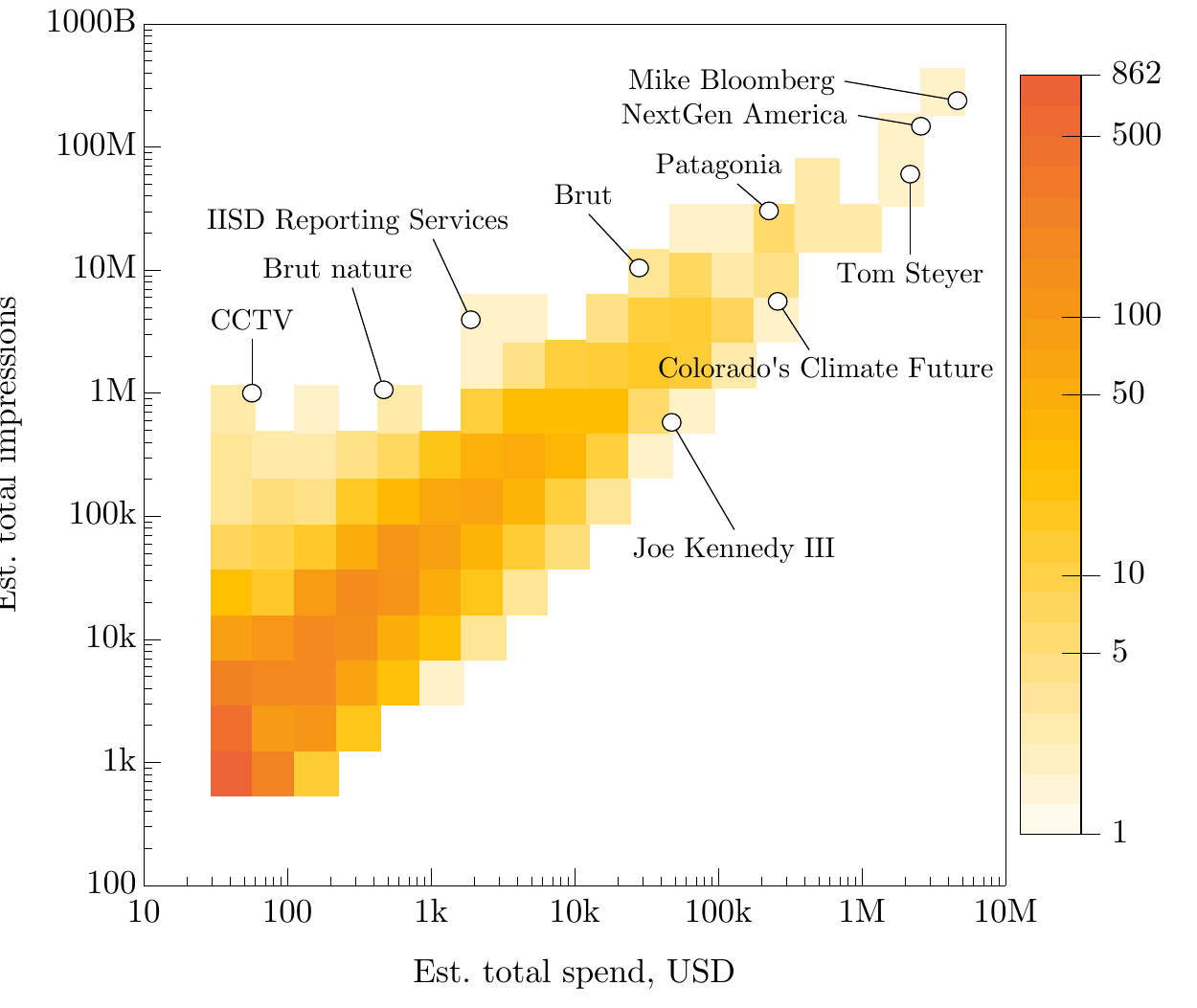}\label{fig:spend-impres}}
    \subfloat[][]{ \includegraphics[width=.49\textwidth]{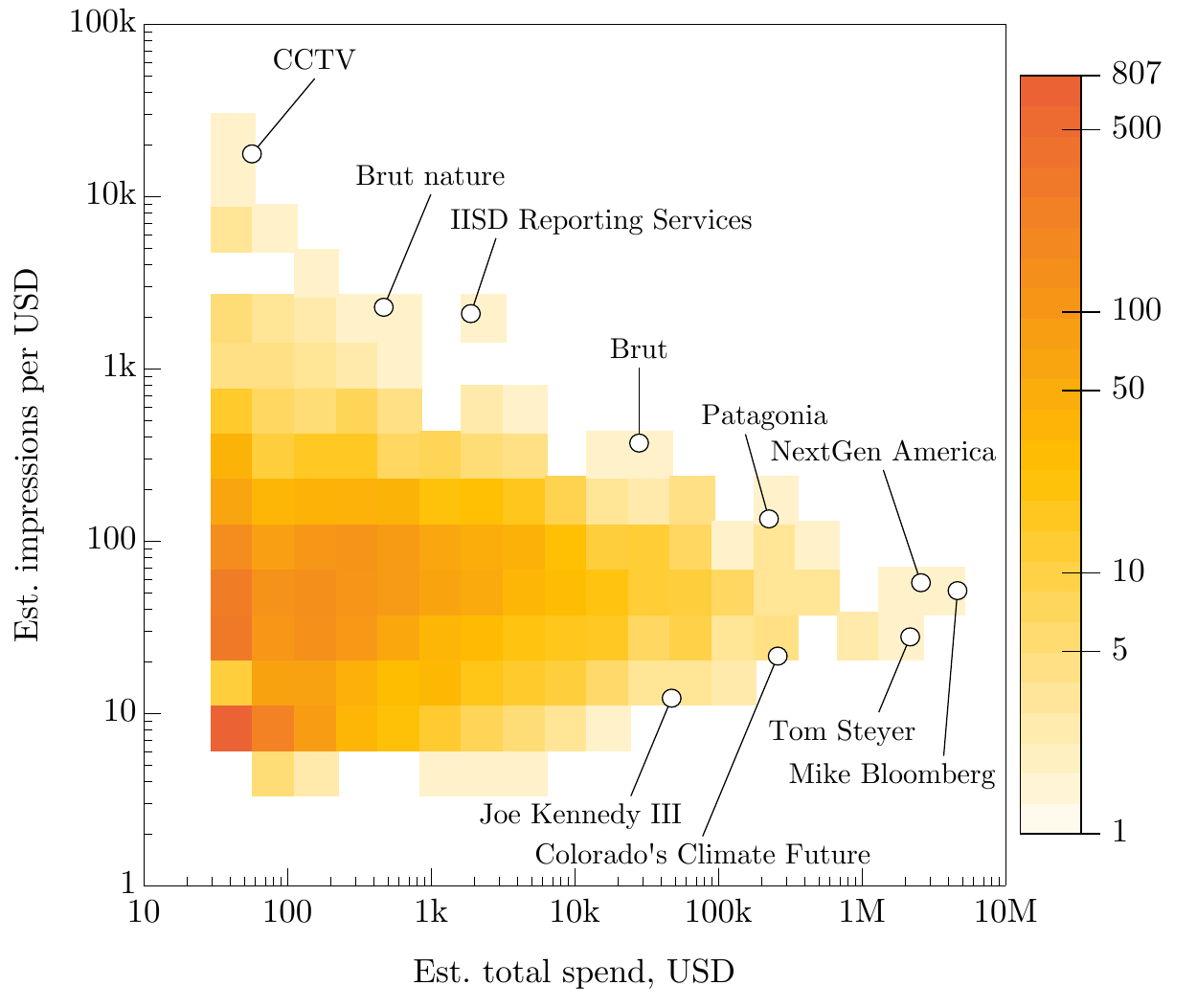}\label{fig:impres-per-spend}}
    \caption{Spend and impressions for sponsored Facebook posts related to climate change in the United States. Both panels show heatmap density distributions of the number of posts with different characteristics. 
    Figure \ref{fig:spend-impres}: Impressions vs spend for posts sponsored by different Facebook Pages. Illustrative example••••••••s of sponsors with different spend-vs-impressions profiles are highlighted. There is a strong positive relationship between spend and impressions. Figure \ref{fig:impres-per-spend}: Impressions per USD vs total spend for posts sponsored by different Facebook Pages. Illustrative examples of sponsors with high/low returns within different total spend bands are highlighted. The number of impressions per USD can vary by several orders of magnitude, especially in the lower total spend bands.}
    \label{fig:impres-spend}
\end{figure}

Figure \ref{fig:impres-spend} shows the relationship between spend and impressions as heatmap density distributions of the number of sponsors falling into different spend-vs-impressions bands. Distributions show posts aggregated by sponsor; recall that the sponsor of a post is identified by the Facebook Ad Library API as a Facebook Page, which may belong to an organisation or an individual. Figure \ref{fig:spend-impres} therefore shows the distribution of total spend and total impressions by different sponsoring Pages, with a number of exemplar Pages highlighted that have significantly high or low impression volumes for their spend level. 

Figure \ref{fig:impres-spend} shows that most sponsors have relatively low spend and low impressions. There is a strong positive association between total spend and total impressions; higher spend results in more impressions, as might be expected given that these adverts are paid for by sponsors. However, there is wide variation (up to three orders of magnitude) in the number of impressions within the same spend band, especially for lower spend bands. So while post sponsors get (at least) `what they pay for', there is a huge potential benefit in additional impressions from cost-free social sharing. This is recapitulated in Figure \ref{fig:impres-per-spend}, which shows the distribution of sponsors by number of impressions per USD of spend (i.e. the number of total impressions divided by the total Page spend). Interestingly, here there appears to be a negative correlation between total post sponsorship spend and impressions per USD, suggesting that big spenders gain relatively fewer socially shared impressions. Figure \ref{fig:impres-per-spend} illustrates this, where Pages with lower total expenditure generally enjoy a greater amount of exposure for the money they spend.
Figure \ref{fig:impres-spend} also highlights some sponsors that exemplify the variation in advert impressions achieved per unit spend.

\section{Results}
\label{results}

Having looked at the `where', `when' and `how much' of sponsored posts relating to climate change on Facebook during the Exploratory Data Analysis section, we next turn to the `who' and `what' in our main Results. In this section, we examine the main types of actor that sponsor climate-related paid content on Facebook (Section \ref{sec:sponsors}), the content of the sponsored posts separated into visual imagery (Section \ref{sec:postcontent-imagery}) and text (Section \ref{sec:postcontent-text}), and how content is used by different actor types to frame the issue of climate change (Section \ref{sec:postframing}). 

\subsection{Post sponsors}
\label{sec:sponsors}

In this section, we explore the top-ranked sponsors of climate advertising by spend and impressions, seeking to characterise the types of actor that are most prominent in sponsoring Facebook posts around climate change. The most prominent Facebook Pages sponsoring posts for each country were manually labelled into four categories: environmental non-governmental organisations (ENGOs), other non-governmental organisations (NGOs), political parties and/or political candidates, and commercial entities promoting particular products or services. 

\begin{figure}
    \begin{subfigure}{1\textwidth}
        \centering
        \includegraphics[height=8cm]{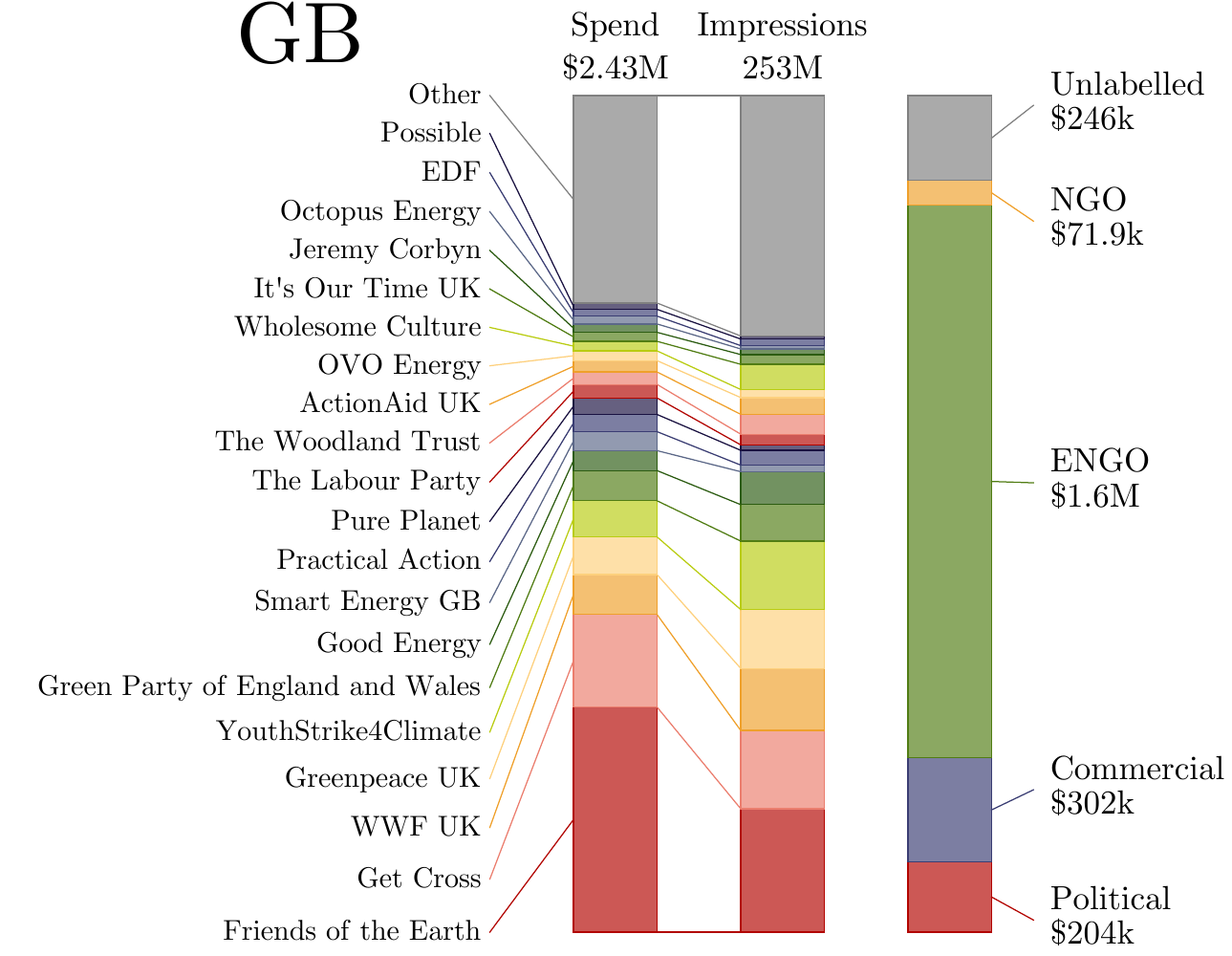}
    \end{subfigure}
    \\
    \vspace{5mm}
    \begin{subfigure}{1\textwidth}
        \centering        
        \includegraphics[height=8cm]{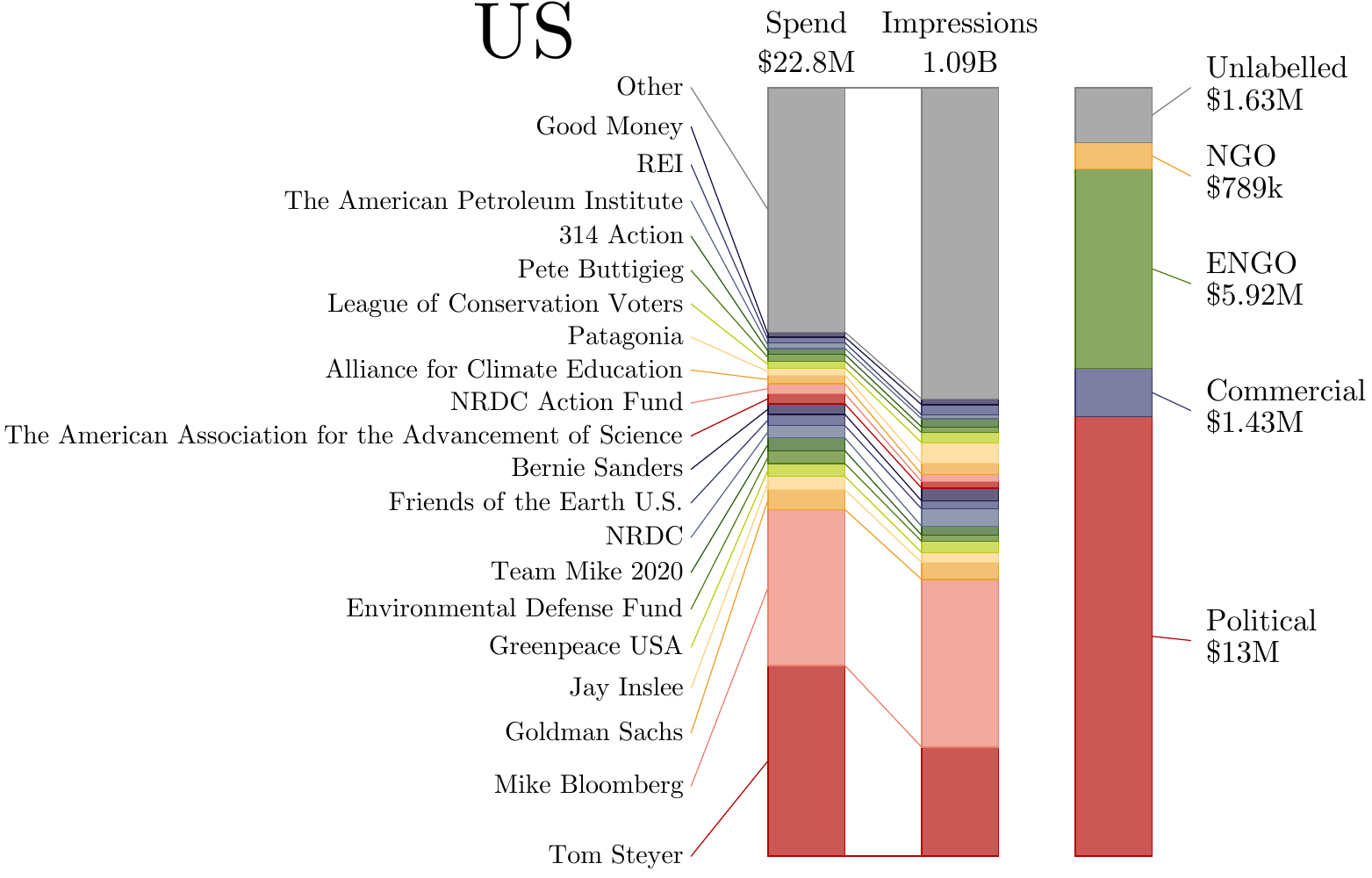}
    \end{subfigure}
    \caption{Top-20 sponsors by total spend value on climate-related Facebook posts in US and GB. In each panel, two linked stacked bar charts (left) show the proportion of total spend and impressions achieved by top-20 sponsors in each country. A separate stacked bar chart (right) shows the share of total impressions for different categories of sponsor (political, commercial, ENGO, NGO) within the top-20 sponsors from each country, with share of total spend given in the labels.}
    \label{fig:sponsors-US-GB}
\end{figure}

Figure \ref{fig:sponsors-US-GB} shows the top-20 sponsors of climate-related posts in US and GB; similar data for CA are given in Figure S3. Sponsors are ranked by their proportion of total spend in that country, as well as their share of total post impressions. The top-20 sponsors are different in each country, with the only overlap between countries the appearance of Friends of the Earth and Greenpeace in the top-20 lists for both US and GB (albeit different geographic divisions of those organisations). There are some recurring patterns for all countries. Political parties and election candidates are prominent (e.g. Mike Bloomberg, Tom Steyer and Jay Inslee all stood for nomination as the Democrat party presidential candidate in the US; Green Party and Labour Party are visible in the UK; Liberal Party and Justin Trudeau are visible for Canada). Environmental non-government organisations (ENGOs) are also prominent sponsors of posts (e.g. NextGen America is a progressive campaigning foundation funded by Tom Steyer in the US; Woodland Trust, Greenpeace and WWF are environmental charities popular in GB; Fair Path Forward is a Canadian non-profit organisation that advocates for pro-environmental policies). More specific campaigns and action groups are also visible (e.g. Get Cross and YouthStrike4Climate in GB; Boreal Conservation in CA; Alliance for Climate Education in the US). Commercial actors are also present, typically representing brands appealing to ecological/environmental values or seeking to do so (e.g. Goldman Sachs and Patagonia in US; Good Energy, Patagonia and Volvo in UK; Ethical Bean Coffee and Unilever in Canada).

As to be expected given the previous analysis of spend and impressions, there is huge variability between sponsors in their spend/impressions. Figure \ref{fig:sponsors-US-GB} and Figure S3 show that in all three countries a few big sponsors are responsible for a large fraction of spend and impressions, with a larger group of smaller sponsors responsible for the remainder. The US has 5,488 different sponsors in total, of which the top-20 by spend produced 60\% of the total spend and 52\% of the total impressions. GB has 1,259 different sponsors, with the top-20 producing 74\% of spend and 71\% of impressions. CA has 832 sponsors with the top-20 producing 65\% of spend and 59\% of impressions. 

Consistent with the wide variation in volumes over time for sponsored posts (see Figure \ref{fig:concurrent_adverts}), individual sponsors are also highly variable in when they place posts (data not shown). Most strikingly, US presidential candidates Mike Bloomberg and Tom Steyer were the highest spending Facebook Pages in the US dataset; Bloomberg published posts exclusively between January and April 2020, outspending most other Pages in a matter of months. (Exactly which of these two actors has spent the most in this dataset depends on estimates of the lowest spend-band value, with Steyer's Page producing many more low-spend posts resulting in lower certainty.) 

Figure \ref{fig:sponsors-US-GB} also shows the total aggregated spend by four categories of actor (based on examination of the sponsoring Facebook Pages): ENGOs, NGOs, political, commercial. The total spend by actors in each category varies between countries. ENGOs dominate spend in the GB dataset, while political actors are somewhat more prevalent in CA and much more prevalent in the US dataset. This may reflect the timing of the study period, which captures part of the US presidential election process, or a more general feature of climate-related Facebook advertising. In all countries, there is a smaller but significant spend by commercial entities. These actors typically promote a certain product or service using an environmental theme (e.g. conservation-themed apparel, climate-conscious investment plans). 

\subsection{Post content: visual imagery}
\label{sec:postcontent-imagery}


Visual content (images, video) appears in almost all sponsored posts and is perhaps the most striking feature of a post. Posts can feature anything from a single visual item to half a dozen, which may be still images or short videos. Here we do not analyse video content in its natural form, but instead use the ``poster'' frame (the still image which appears in a browser until the video is played, typically the first frame of the video); this is similar methodologically to newspaper analyses that focus on the lead or header image. To explore the images used in climate-related sponsored posts, we first create image collages to give a holistic impression of how imagery appears and differs between countries, before then performing a more rigorous image categorisation. Details of the methods underlying these analyses are given in Section~S1.4).

\begin{figure}
    \begin{subfigure}{1.0\textwidth}
        \centering
        \includegraphics[width=1\textwidth]{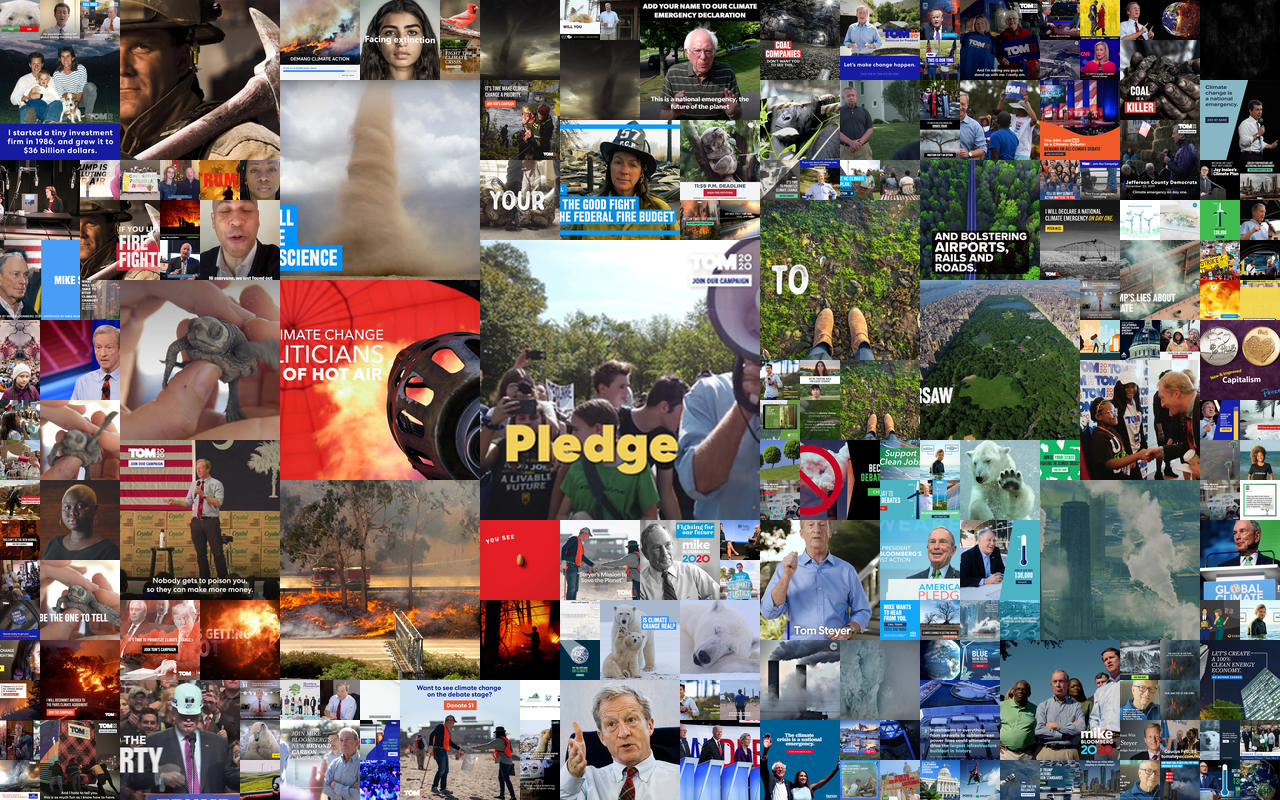}
        \caption{US}
    \end{subfigure}
    \\
    \begin{subfigure}{1.0\textwidth}
        \centering
        \includegraphics[width=1\textwidth]{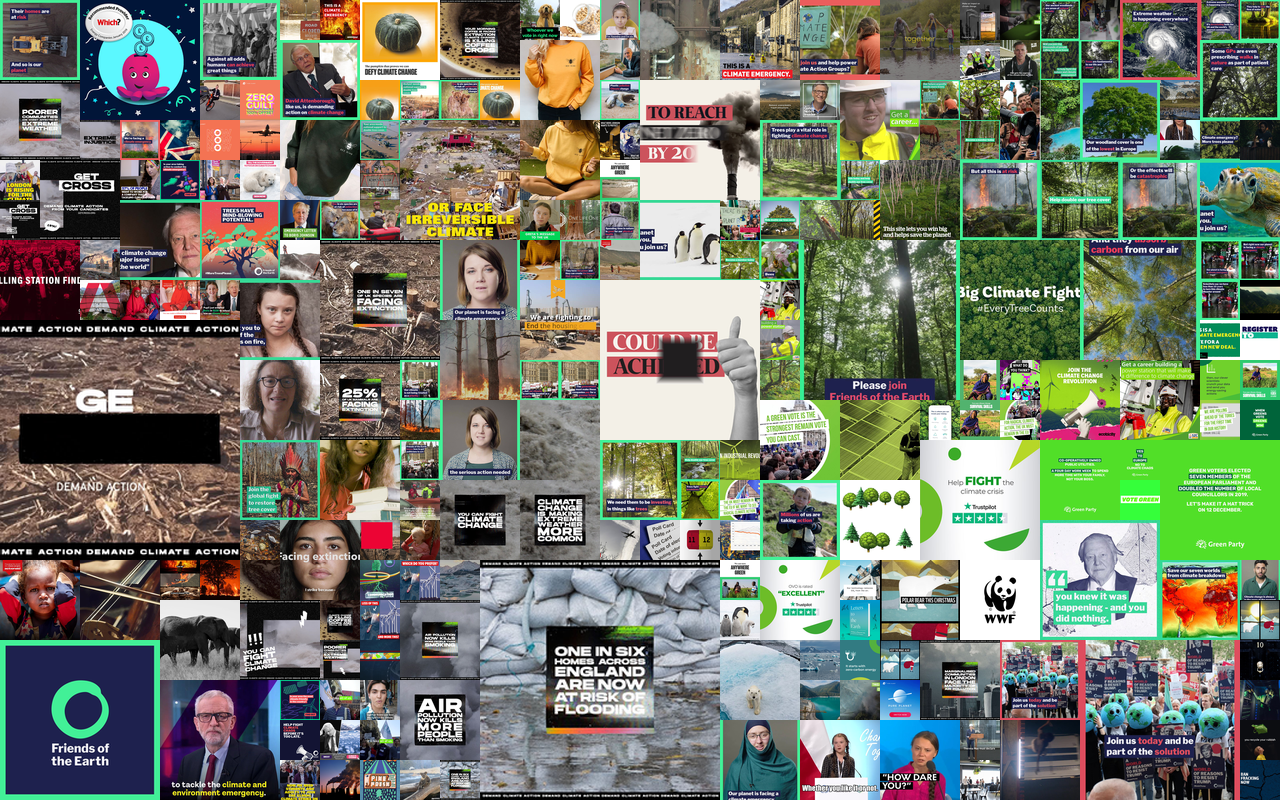}
        \caption{GB}
    \end{subfigure}
    \caption{Collages of sponsored post imagery from US and GB datasets. The top-ranked images by total spend are shown (approximately 200 images per country). Image size reflects the total spend value on posts incorporating that image. Manual inspection reveals common themes (e.g. prominence of people and text in many images) and differences (e.g. localised political figures) between countries.}
    \label{fig:collages}
\end{figure}

 Figure~\ref{fig:collages} shows a collage of approximately 200 top-ranked images by total post spend value for US and GB; a similar plot is given for Canada in Figure~S4. High-resolution versions of these are hosted at \url{https://photos.app.goo.gl/ctrL7QFTyyA4Ddn49}. Collages were created using image hashing to group visually similar images in the frame, with the displayed size of each image determined by relative spend; where the same image was used in multiple posts (a common practice), we aggregate the total spend on adverts using that image. 
 
 The collages provide a holistic impression of the kinds of imagery used in climate-related sponsored posts in each country. One general observation is that most images in climate advertising involve people, a finding seen elsewhere for most news media coverage  \citep{oneill2013imagematters, difrancescoyoung2011visualconstruction}. Looking at the people in the images, there is some localisation with nationally important figures visible for each country. Strikingly, in the US and CA sponsored posts often portray political candidates and leaders, typically displayed in a positive way (e.g. with good lighting, smiling). By contrast, political figures are rare in the GB dataset; right-wing politician Nigel Farage appears rendered satirically as a vampire, with similar negative tropes applied to images of other right-wing politicians. Some internationally recognised individuals are seen in collages for all three countries, such as climate activist Greta Thunberg. Another common observation is images with superimposed text, widely used in all three countries. Also common are images of landscapes and nature, though there are geographic  differences; the US dataset shows a lot of climate impact imagery, such as extreme weather (e.g. storms, wildfires) whereas GB and CA show less emotionally loaded natural imagery (e.g. trees, mountains). Polar bears, an iconic visual metonym for climate change \citep{oneill2022bears}, can be seen in posts delivered to all three countries.

\begin{figure}
    \begin{tabularx}{1.0\textwidth}{l l l}
    \multirow{9}{0.8\textwidth}{\includegraphics[width=.75\textwidth]{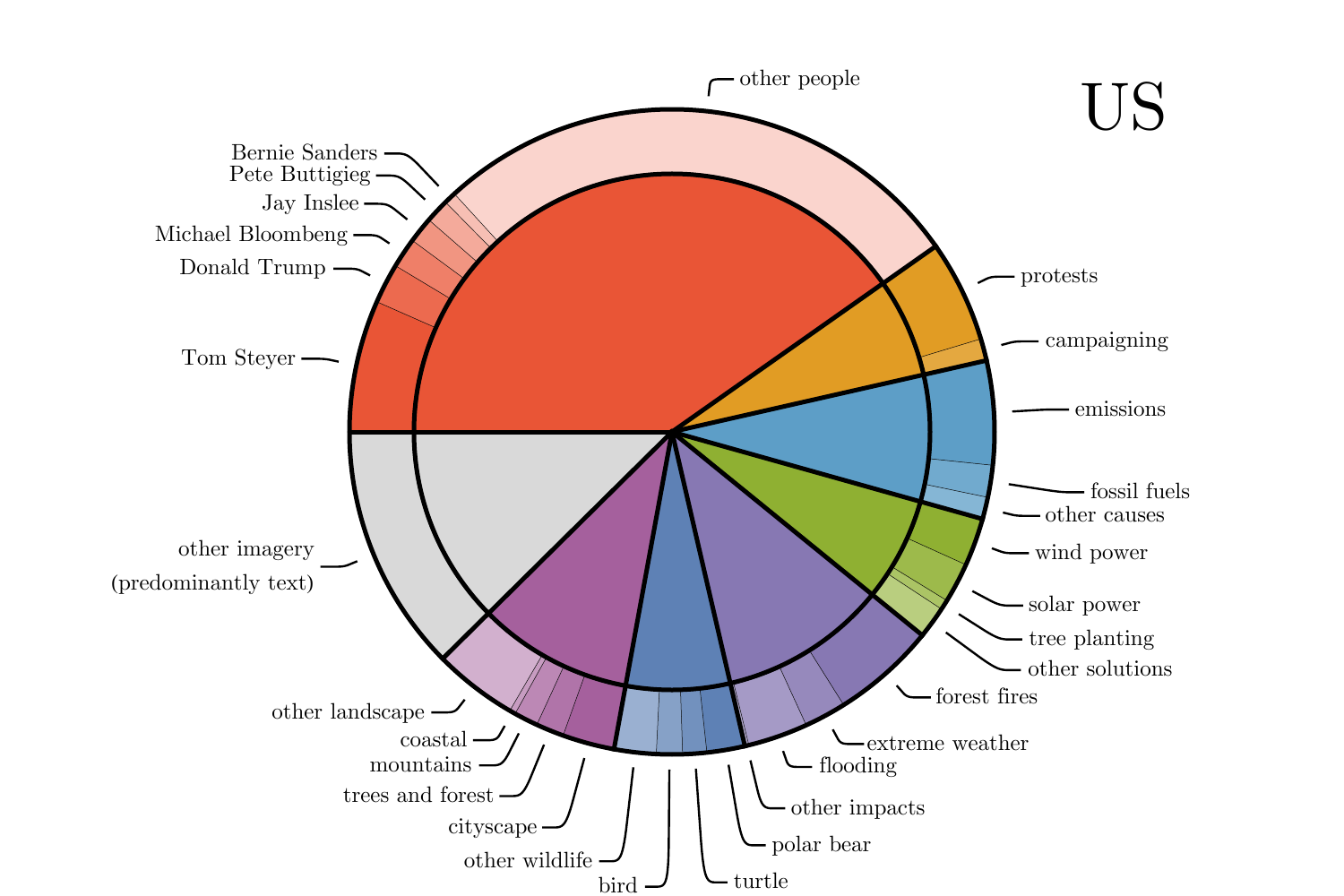}}
    & \fcolorbox[rgb]{0,0,0}{0.915,0.3325,0.2125}{\rule{0pt}{6pt}\rule{6pt}{0pt}} & People\\
    & \fcolorbox[rgb]{0,0,0}{0.772079,0.431554,0.102387}{\rule{0pt}{6pt}\rule{6pt}{0pt}} & Collective action\\
    & \fcolorbox[rgb]{0,0,0}{0.363898,0.618501,0.782349}{\rule{0pt}{6pt}\rule{6pt}{0pt}} & Causes\\
    & \fcolorbox[rgb]{0,0,0}{0.560181,0.691569,0.194885}{\rule{0pt}{6pt}\rule{6pt}{0pt}} & Solutions\\
    & \fcolorbox[rgb]{0,0,0}{0.528488,0.470624,0.701351}{\rule{0pt}{6pt}\rule{6pt}{0pt}} & Impacts\\
    & \fcolorbox[rgb]{0,0,0}{0.368417,0.506779,0.709798}{\rule{0pt}{6pt}\rule{6pt}{0pt}} & Wildlife\\
    & \fcolorbox[rgb]{0,0,0}{0.647624,0.37816,0.614037}{\rule{0pt}{6pt}\rule{6pt}{0pt}} & Landscape\\
    & \fcolorbox[rgb]{0,0,0}{0.85,0.85,0.85}{\rule{0pt}{6pt}\rule{6pt}{0pt}} & Other imagery\\
    & & (predominantly text)\\[2cm]
    \includegraphics[width=.75\textwidth]{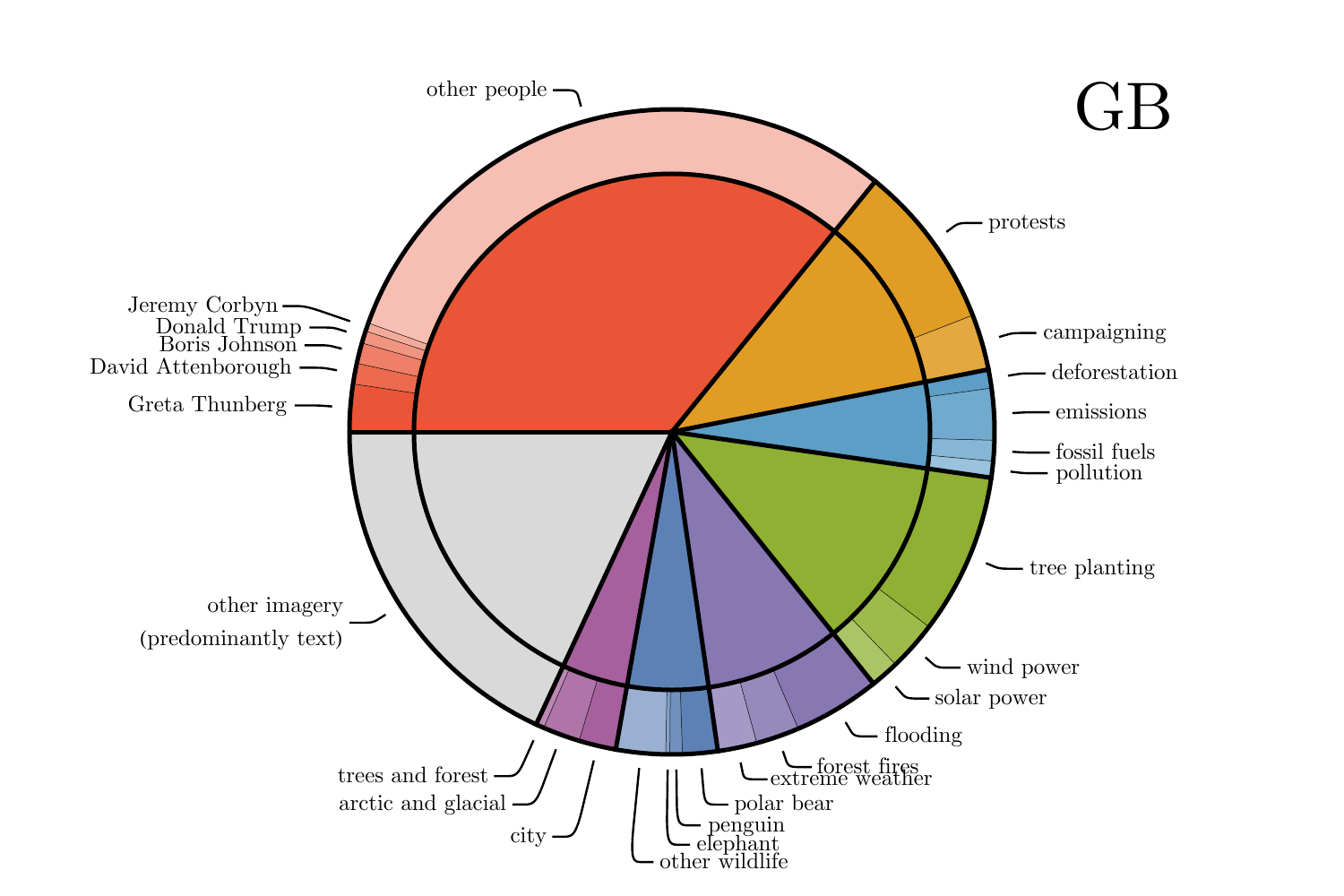} & \\
    \end{tabularx}    
    \caption{Categorisation of images used in climate-related Facebook posts in US and GB. The top-2000 images (by total spend on adverts using each image) were manually classified in a scheme using eight main categories. Popular sub-categories are indicated in a second-level categorisation within each colour range.}
    \label{fig:imagery}
\end{figure}

We manually tagged images into eight broad categories (see Section~S1.4.1 for coding scheme): `people', `collective action', `causes', `solutions', `impacts', `wildlife', `landscape', `other imagery'. Within each broad category, we also labelled sub-categories when there were many images of the same thing (e.g. the same individual person, a particular type of landscape). Figure ~\ref{fig:imagery} shows the most abundant image categories across the top-2000 images for the US and GB; since the manual tagging was very time-consuming, we did not categorise images from Canada. For each plot, the top-2000 images were identified by the total spend on adverts including that image, i.e. each image may have appeared in more than one advert. 

We find that `people' is the most-used image category for both US and GB, as suggested by the collages. 
US posts are populated by a diversity of politicians, especially candidates that were seeking nomination for election as President during the study period. For the UK, the internationally recognised climate activist Greta Thunberg is most often used, followed by David Attenborough, the popular presenter of wildlife television documentaries, followed by a number of political leaders. Meanwhile, the `collective action' category is also focused on people and well-represented in both US and GB climate adverts. These images are differentiated into `protest' imagery (e.g. pictures of marches and public gatherings) and `campaigning' imagery (e.g. political rallies). Together the human-focused categories of `people' and `collective action' cover almost half of the top-ranked advert images for both the US and the UK.

The `causes', `solutions' and `impacts' of climate change together make up around a quarter of sponsored post images in both the US and the UK, though their proportions differ between the two countries (`impact' and `cause' images are more common in US, while `solution' images are more common for GB). `Cause' imagery tends to focus on emissions (often visualised with smoke stacks) and fossil fuels (particularly imagery of less conventional, more controversial, sources such as tar sands and fracking). `Solution' imagery shows renewable energy sources and, for the UK, tree-planting. `Impact' imagery is quite localised, with an emphasis on forest fires in the US and floods in the UK, reflecting extreme weather types experienced in each country during the study period. 

Of the remaining categories, `wildlife' imagery is a minor but noticeable category for both the US and UK, while `landscapes' are more commonly seen in the US. In the `other imagery' category, a common phenomenon is images that consist only of text (that is, a picture of some text, perhaps in a bold or unusual font). This is widely seen in all countries studied.

\subsection{Post content: text}
\label{sec:postcontent-text}

\begin{figure}
    \includegraphics[width=.75\textwidth]{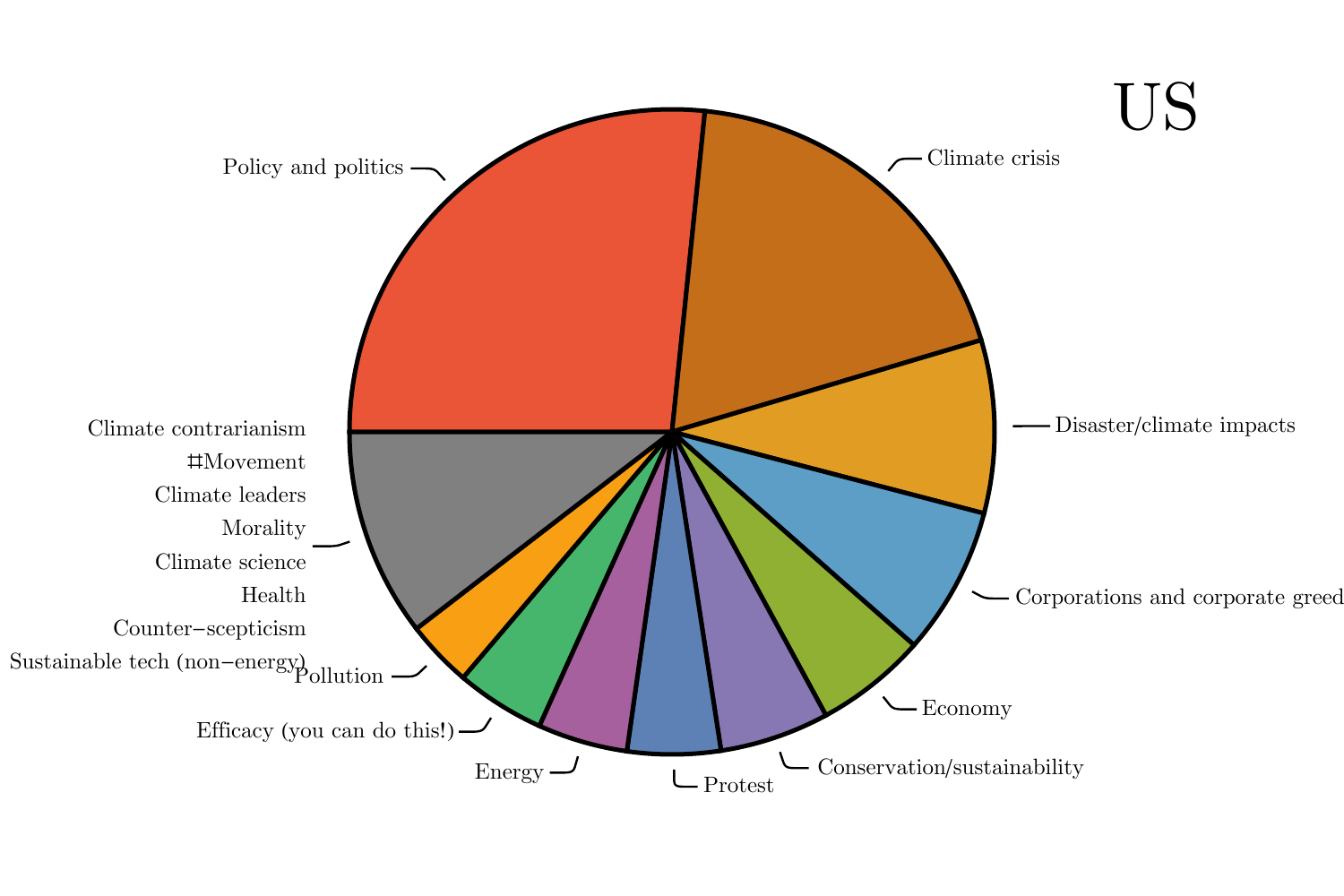}\\
    \includegraphics[width=.75\textwidth]{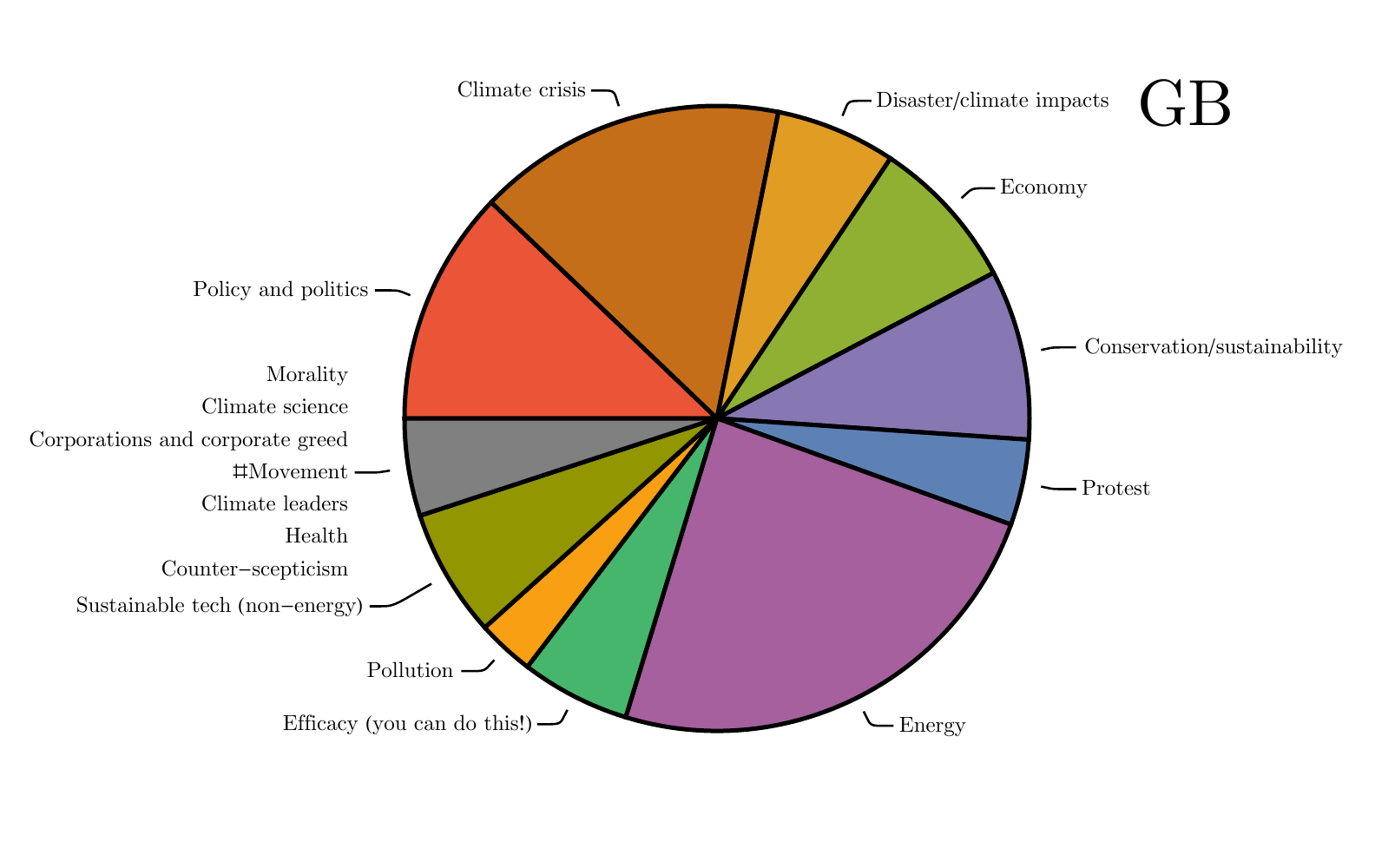}
    \caption{Categorisation of text content in climate-related Facebook posts in US and GB. Text from adverts containing the top-2000 images (by total spend on adverts using each image) was manually classified in a scheme using eighteen main categories. There are a number of small categories which would not be individually visible in the charts, so are grouped into a miscellaneous wedge for each country.}
    \label{fig:text}
\end{figure}

For the same set of adverts that provided the images categorised in the previous section (i.e. the adverts containing the top-2000 images by spend), we manually categorised the text content into 18 broad categories (see Section~S1.5 for coding scheme). The diversity of categories reflects the complexity of modern climate change discourse, with visible categories including: `policy and politics', `climate crisis', `corporations and corporate greed', `disaster/climate impacts', `economy', `protest', `efficacy (you can do this!)', `energy', `conservation/sustainability', `pollution', `\#Movement', `climate leaders', `climate contrarianism', `climate science', `morality', `health', `counter-scepticism', `sustainable tech (non-energy)'. Figure ~\ref{fig:text} shows the most abundant text content categories across the top-2000 adverts for the US and GB; again, since the manual tagging was time-consuming, we did not categorise text for Canada.

Figure ~\ref{fig:text} shows a few interesting patterns in how text is used in sponsored posts about climate change. Many adverts focus on `policy and politics', which is the largest category in the US (reflecting the ongoing Presidential primary campaign ongoing during the study period) and also a substantial category in the UK. Advert texts referencing or drawing attention to the `climate crisis' are prominent in both the US and the UK. The US has a substantial category of adverts that talk about `corporations and corporate greed' in the context of climate, which is rarely seen in the UK. The UK has a large number of adverts focused on `energy', which is much less frequent in the US. There are also a number of small but noteworthy categories, including `climate contrarianism' which is mostly seen in the US, and its ideological converse, `counter-scepticism'.

\subsection{Framing of climate change by different actors}
\label{sec:postframing}

\begin{figure}
    \centering
    \includegraphics[width=1\textwidth]{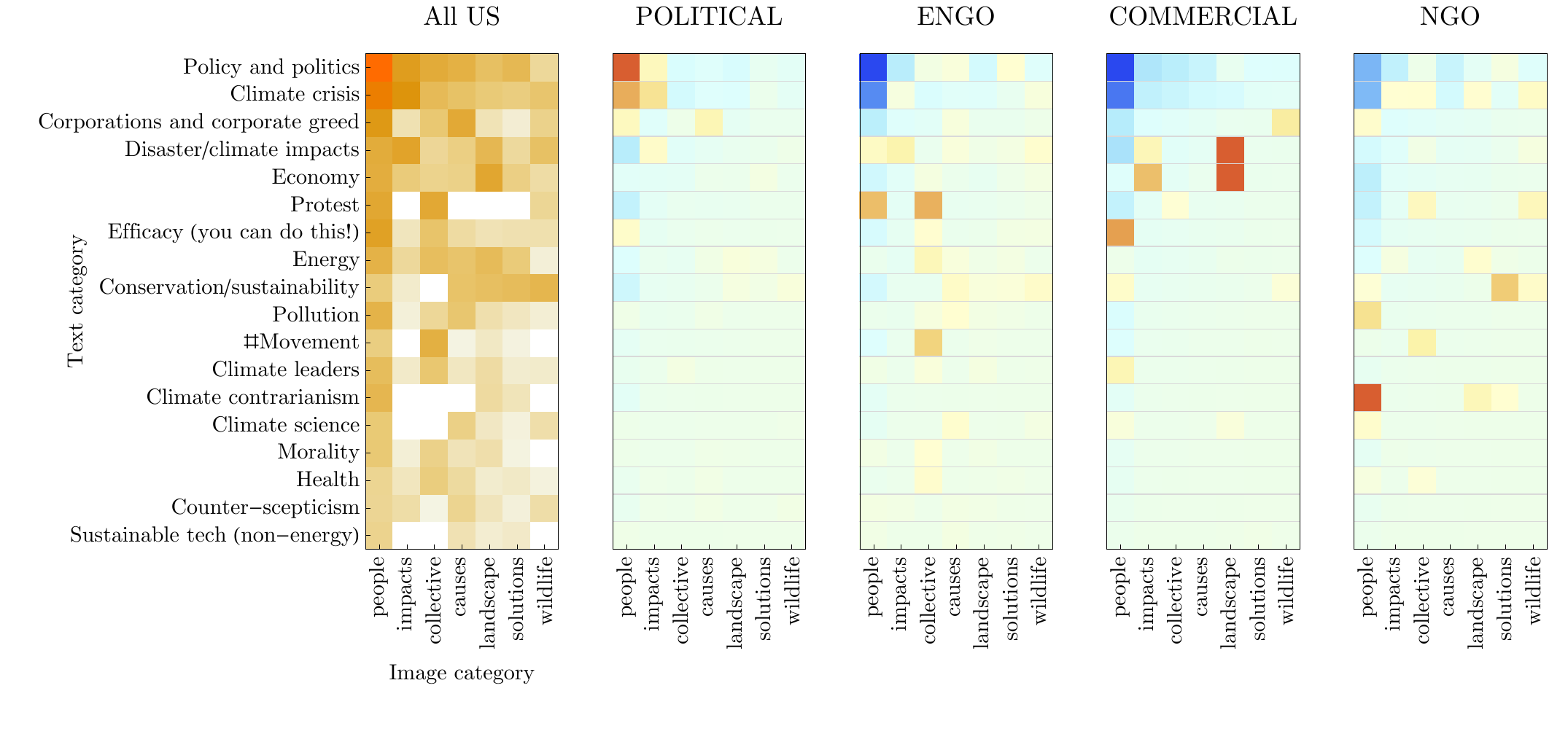}
    \\
    \includegraphics[width=1\textwidth]{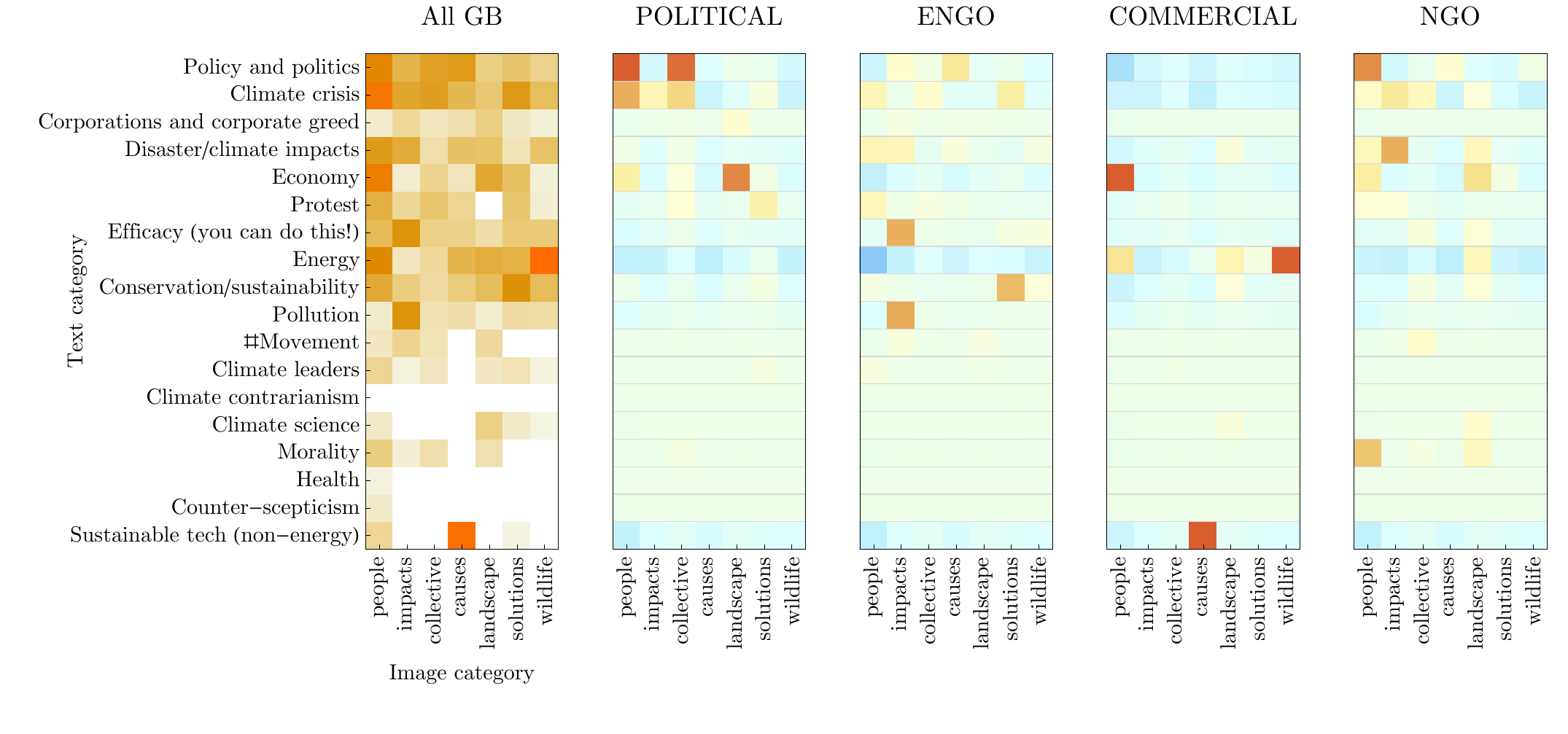}
    \caption{Combination of image and text categories in climate-related sponsored posts on Facebook for the US (top) and UK (bottom). The left-most panel shows the frequency of adverts that use each pairwise combination of image and text categories across all of the adverts containing the top-2000 images by total spend. The other panels show how the usage of each image/text combination by a particular actor type varies from the background level (i.e. the deviation from the overall usage), shown as a red (increased usage) to blue (decreased usage) colour scale.}
    \label{fig:textImgComp}
\end{figure}

Text and imagery are typically combined within a Facebook advert to convey an overall message and frame the issue of climate change in a particular way. In this section, we look at how different actor types (political, commercial, environmental and non-environmental NGOs) combine images and text in their sponsored posts, and what this suggests about how they frame the issue. Figure \ref{fig:textImgComp} shows the frequency of adverts in the US and UK that show different combinations of image and text categories, using the manually coded category labels generated in the previous sections.

The overall frequency of different image/text combinations (Figure \ref{fig:textImgComp}, left panel) to some extent reflects the overall abundance of the underlying image and text categories, with a high abundance of `policy and politics' text and `people' imagery. This creates a high usage of these dominant categories in combination with all other categories of the other type. Beyond that, a number of patterns can be observed. In the US and GB, text relating to `conservation/sustainability' is often paired with imagery related to `landscape' and `wildlife', as well the `causes' and `solutions' for climate change. In the UK, text relating to `energy' is often paired with images of `wildlife', but this pattern is not seen in the US. In the US, `collective action' images are often used with `\#Movement' or `protest' text content, but this is not seen in the UK. In the US, text on `corporations and corporate greed' is often linked to `causes' of climate change, but in the UK there is less of this text in general. In the UK, but not in the US, images of `impacts' are often linked to text about `pollution'. 

Looking across the different actor types (Figure ~\ref{fig:textImgComp}, right panels), it is clear that political actors (either individual candidates or political parties) very commonly use text content related to `policy and politics' and mostly associate it with imagery of `people' (in the US and UK) and also with `collective action' (in the UK only). This usage appears to frame climate change as a political issue focused on humans and political decision-making; this framing also renders the issue the subject of electoral choice and differentiation between candidates/parties. Political actors in both UK and US also use text related to the `climate crisis' and imagery related to climate `impacts', perhaps suggesting a framing of the issue as urgent and salient to the public. In the UK only, there is a high frequency of sponsored content by political actors that combines imagery of `landscape' with text about `economy'; here the landscape and natural imagery may simply reflect the difficulty of finding visual imagery to represent the human construct of the `economy', but the use of economic framing suggests a focus on this aspect in the UK. 

Environmental NGOs show different framing in the US and the UK. In the US, there is increased usage of text relating to `protest' with imagery associated with `collective action', such as public gatherings and marches. This suggests a framing of climate change as something which can be tackled by collective behaviour and grassroots support for action, or a reason to be recruited to a cause, or a motivation for the public to call for change. In the US, ENGOs also use `\#Movement' text alongside `protest' imagery, perhaps indicative of online social media campaigns and viral engagement. However, in the UK, ENGOs tend to associate imagery of `impacts' with text about `pollution' and `efficacy', perhaps suggestive of campaigns to motivate people to take action in this specific area. 

For commercial actors, the patterns again differ between the US and the UK. In the US, commercial actors often combine imagery of `people' with `efficacy (you can do this!)' text content to frame the issue as one where individual action is important; presumably, the individual action that is suggested is to buy a product or service. Commercial actors in the US also use `economy' text with `impacts' imagery to frame climate change as a threat to economic prosperity, and generic `landscape' imagery with text around `economy' and `disaster/impacts', suggesting a similar framing of climate change as a risk to the economy. In the UK, commercial actors combine `economy' text with imagery of `people', `energy' text with images of `wildlife', and `sustainable tech' text with imagery of `causes'; it is hard to interpret the purpose or framing here without further manual inspection, but some examples are suggestive of positioning the commercial entity as a sustainable or green option within the market, perhaps as a point of differentiation from competitors. 

An interesting combination used by non-environmental NGOs is the association of `people' imagery with `climate contrarianism' text content in the US; this is not seen in the UK and is suggestive of a counter-framing of the climate change issue as something that people should be unconcerned or doubtful about, consistent with the contrarian agenda of denial and delay. Meanwhile, UK posts by non-environmental NGOs connect the text theme of `morality' with images of `people', perhaps appealing to personal conscience to promote climate-related actions.

\section{Discussion}
\label{discussion}

In this study, we have conducted the first systematic exploration of sponsored posts (adverts) related to climate change on the Facebook platform, using a large dataset from the Facebook Ad Library service. Data was collected for a 22-month period during 2018-2020 for the English-speaking countries of United States (US), United Kingdom (UK) and Canada. Several analyses of this data are presented to give an empirical understanding of what kinds of advert are seen by users, in what volumes, from which actors, and how the sponsored posts frame the issue of climate change.

In the data we studied, total spending on climate-related adverts is largest in the USA (12.3M USD in the comparison period), UK (2.87M USD) and Canada (1.32M USD), although per capita spending is relatively similar (USA: 0.04 USD, UK: 0.04 USD, Canada: 0.03 USD). This suggests that there is strong interest in sponsored content around climate change in all three countries, which represent advertising markets of different sizes. The volume of advertising varies considerably over time, with distinct peaks of activity that are mostly associated with political events in each country. The main relationship between spend and reach of climate-related adverts is a strong positive correlation (i.e. total reach is primarily driven by total spend), with a secondary effect of social sharing whereby some advert sponsors get greater reach than others for similar spend. Different types of actor dominate the total spend in each country, with political actors dominant in the US and environmental NGOs dominant in the UK.

Looking at content, most climate-related sponsored posts contain images of people, either individuals or groups, with smaller but significant usage of imagery related to the causes, impacts and solutions for climate change, as well as natural imagery of wildlife and landscapes. Text content showed a wide variety of frames and narratives, including political arguments (especially in the US), framing of the issue as a `climate crisis' (in both US and UK) and a thematic focus on energy in the UK. Combining text with imagery, we performed a simple analysis of how different actor types frame the issue of climate change. Political actors focused on climate as a human issue requiring a political solution, with text related to policies and images of people and collective action. Environmental NGOs frame climate change as something where personal and collective action is important, emphasising the climate crisis, protest and personal efficacy with pictures of people, climate impacts and climate solutions. Commercial actors in the US focused on economic aspects and climate impacts, but in the UK focused on energy and sustainable technologies. Non-environmental NGOs in the US were notable for a focus on climate contrarian/sceptic views, which was absent in the UK.

There are several limitations to this study. Firstly, the dataset was collected from a `black box' service provided by Facebook Ad Library. Known limitations of this service are that it only provides adverts/posts that have been tagged as `social or political' in purpose, but the methods for this tagging are opaque, as are the completeness and coverage of the sample that is returned. Thus our dataset does not include private user content from the general population or commercial content without a social/political purpose (i.e. most products and services). Our sampling strategy also limits the scope of the study by focusing on a particular time period, on English-language content (via our use of English-language keywords) and our analysis on a particular geography within the sample (US, UK, Canada). Our use of the `climate' keyword to collect data may have excluded some relevant content (e.g. discussion of renewable energy not referencing climate change). However, we do not believe that these sampling limitations have substantively affected the study. Perhaps the main exclusion in the data is content that uses climate change as a way of selling commercial products or services. While we do see some commercial actors and a number of product advertisements (e.g. `Ethical Bean Coffee' in Canada), it is reasonable to assume that there is a much larger set of commercial advertising that uses climate to draw attention or position a product (perhaps disingenuously, cf. `greenwash'), which is not included in this study due to limitations of the Facebook Ad Library. Regarding the time period of the study, we first note that longitudinal studies of media content are rare in the academic literature. Here we present a 22-month period that includes a number of key political and other events. Our geographic focus on US, UK and Canada is reflective of the language spoken by the authors and we note that the Facebook Ad Library makes available data for a number of other (non-English-speaking) geographies which would form an interesting target for future study. One additional limitation relates to our methodological approach, where we tried to strike a compromise between scale (covering large datasets from three countries) and depth (performing human annotation of images and frame-coding of text). 

One finding of our study concerns the multimodal nature of the data. The sponsored posts always contain text and almost always contain images and/or video content. In our analysis, we handled this by considering each modality separately, manually categorising the images and text content, then associating the categories across the modalities based on the frequency of each combination. While this approach is reasonable, we speculate that future studies of sponsored Facebook content (and other online media content) might usefully combine the modalities and understand the content as it is presented to the users of the platform; that is, as a multimodal, integrated, package of text and video/imagery. Holistic interpretation of post content might better reveal how text and imagery are combined to present the issue of climate change. A further obvious step is to consider video content; here we used only a single still image (the so-called `poster' image) to represent embedded video content. This was necessary to handle the volume of data in our study, but an aspect that might be improved in future work. The rise in video-based social media services (e.g. YouTube, TikTok) shows an increasing usage of video content in online discourse and we believe that it is an important subject for academic study, notwithstanding the methodological challenges of doing so at scale.

Contrasting our findings for different countries, we find that climate-related sponsored content is more political in the US than in the UK, with Canada somewhere between. This possibly reflects the study period, which saw a greater number of political events in the US than in the other two countries (e.g. the mid-term elections and presidential primary campaigns in the US vs the general election in the UK). However, the data still appears to show a greater abundance of politically driven sponsored posts on the topic of climate for the US than the two comparator countries. We speculate that this reflects a greater level of politicisation of the issue in the US, where other studies have shown that climate is a party-political issue that has long been both polarised and politicised (e.g. \cite{mccrightdunlap2011, dunlapmccright2016}; see also \cite{fisher2022} on the rise of such politicisation in Europe). In the UK, climate change is more commonly framed as a collective action or movement, with greater activity by environmental NGOs in produced sponsored content.

Another aspect of the politicisation of climate change in the US is the presence of climate sceptic advertising, which was not seen for the UK or Canada. A strong fossil fuel lobby in the US has provided fertile ground for sceptic or contrarian narratives to be maintained. Yet even in the US, the data analysed here shows a low prevalence of sponsored posts that show sceptic/contrarian viewpoints. This is perhaps surprising given the strong presence of such arguments on other forms of social media \citep{williams2015echochambers, coan2021computer, treen2020misinfo, treen2022reddit}. To explore this finding further, we manually examined the Facebook adverts produced during the study period by a list of known climate sceptic thinktanks (including PragerU, Heartland Institute, and others), locating the content by a search on the organisation rather than using keywords. Results showed (data not given) that these groups are active on Facebook and do create a large quantity of sponsored content. However, while they do talk about issues relevant to climate change, they do not tend to talk about climate change directly and importantly, they do not use climate-related keywords that would have introduced them to our data collection. Instead, they talk about counter-point narratives such as job security, energy security, or `creeping socialism'. Understanding the interplay between climate-focused narratives emerging from the environmentalist movement and from their `opposition', as well as how these feed into political arguments, is a key area for future study.

Considering the framing of climate change, we find that the presentation of climate change as an issue varies between actor types, with a general finding that all the actors appear to be using sponsored content on Facebook to promote some kind of action or gather support for a cause. This is perhaps unsurprising given the nature of the data; adverts cost money and are used purposefully. Mass media has traditionally been a ``capital resource" for social movements \citep{tufekci2013attention}, with media platforms acting as gatekeepers between movements and their intended audiences \citep{benfordsnow2000framing}. Here we see widespread activity by a variety of actors that seek to engage large audiences or promote collective action (e.g. `\#Movement' or `efficacy (you can do it!)' frames). This suggests that Facebook (and presumably other social media platforms) are providing a disintermediation mechanism for all kinds of actor, allowing political actors to speak more directly to their audience (thereby adopting new strategies of communication \citep{johansson2019digital}), while NGOs and smaller actors can more easily enlist collective support to challenge or disrupt dominant narratives. Facebook is not limited by (e.g.) the number of pages in a newspaper or the duration of a commercial break, and can place a huge volume of adverts into user feeds with little `editorial' control; indeed there is a vigorous debate about the responsibilities of the platform in this regard \citep{meese_hurcombe_2021}. This presents an opportunity to smaller actors who may not otherwise have been able to reach an audience via mass media. The widespread use of sponsored content in the context of climate change is an example of how movements and advocacy organisations can now ``forcefully offer their framing'' (\citealt{tufekci2013attention}, p. 853) through participatory social media platforms, operating in a shared digital space with media, political parties, and corporate entities \citep{gupta2018advocacygroupmessaging}. This affords them the ability to both circumvent traditional news media and draw journalistic attention to the issues highlighted in their messaging \citep{freelon2020science}. While it is beyond the capacity of our data to infer any effects of advertising on the target populations, the volume of sponsored posts suggests that all kinds of actor in the climate debate find it to be an impactful channel through which to effect their aims.

\section*{Acknowledgements}
\noindent The authors declare no conflicts of interest. This paper was completed with funding from University of Exeter.



\bibliographystyle{elsarticle-harv}
\bibliography{fb_ads.bib}







\end{document}